\documentclass[11pt,oneside,letterpaper]{article}
\usepackage{amssymb}
\usepackage{amsmath}
\usepackage{amsthm}
\usepackage[dvips]{graphicx}
\usepackage{setspace}
\usepackage{amsfonts}
\usepackage{fancyhdr}
\usepackage{xcolor}
\usepackage{graphicx}
\usepackage{rotating}
\usepackage{comment}
\usepackage{color}
\usepackage{cite}
\usepackage{braket}
\usepackage{caption}
\usepackage[utf8]{inputenc}
\usepackage{mathtools}
\usepackage{tikz}
\usepackage{overpic}
\usepackage{stackengine}
\usepackage[normalem]{ulem}
\usepackage{bbold}

\definecolor{darkgreen}{rgb}{0,0.5,0}
\definecolor{darkblue}{rgb}{0,0,0.6}
\definecolor{purple}{rgb}{0.4,.2,0.7}
\newcommand{\p}{\partial}

\newcommand{\be}{\begin{equation}}
\newcommand{\ee}{\end{equation}}

\usepackage[colorlinks=true,citecolor=darkgreen,linkcolor=black,urlcolor=purple]{hyperref}

\usepackage{pdfsync}

\makeatletter
\newcommand*{\defeq}{\mathrel{\rlap{%
                     \raisebox{0.3ex}{$\m@th\cdot$}}%
                     \raisebox{-0.3ex}{$\m@th\cdot$}}%
                     =} 
\makeatother

\DeclareMathOperator{\Tr}{Tr}
\def\be{\begin{eqnarray}}
\def\ee{\end{eqnarray}}

\newcommand{\tr}{\textrm{Tr}\,}

\newcommand{\bea}{\begin{eqnarray}}
\newcommand{\eea}{\end{eqnarray}}
\def\ben{\begin{equation}}
\def\een{\end{equation}}

    \let\p=\phi \let\r=v

 \let\P=\Phi  
\let\C=\Chi 

\def\be{\begin{equation}}
\def\ee{\end{equation}}
\def\ba{\begin{array}}
\def\ea{\end{array}}
\def\del{\nabla}

\def\ba#1\ea{\begin{align}#1\end{align}}
\def\bs#1\es{\begin{split}#1\end{split}}

\renewcommand{\p}{\partial}

\newcommand{\bz}{\bar{z}}

\newcommand{\btau}{\bar{\tau}}

\newcommand{\bh}{\bar{h}}

\renewcommand{\O}{{\cal O}}

\definecolor{vert}{rgb}{0.1367 0.543 0.1367}

\newcommand{\id}{\mathbb{1}}

\allowdisplaybreaks  

\numberwithin{equation}{section}

\def \be {\begin{equation}}
\def \ee {\end{equation}}
	
\def \JM#1 {{\color{blue}  JM: #1 }}
\def \AAl#1 {{\color{red}  AA: #1 }}
\renewcommand{\P}{\mathbf{P}}

\begin{document}
\onehalfspacing

\begin{center}

{\LARGE  {
Toward random tensor networks and holographic codes in CFT
}}

\vskip1cm

Jeevan Chandra and Thomas Hartman

\vskip5mm
Department of Physics, Cornell University, Ithaca, New York, USA

\vskip5mm

{\tt jn539@cornell.edu, hartman@cornell.edu }

\end{center}

\vspace{4mm}

\begin{abstract}

\noindent 
In holographic CFTs satisfying eigenstate thermalization, there is a regime where the operator product expansion can be approximated by a random tensor network. The geometry of the tensor network corresponds to a spatial slice in the holographic dual, with the tensors discretizing the radial direction. In spherically symmetric states in any dimension and more general states in 2d CFT, this leads to a holographic error-correcting code, defined in terms of OPE data, that can be systematically corrected beyond the random tensor approximation. The code is shown to be isometric for light operators outside the horizon, and non-isometric inside, as expected from general arguments about bulk reconstruction. The transition at the horizon occurs due to a subtle breakdown of the Virasoro identity block approximation in states with a complex interior.

 \end{abstract}

\pagebreak
\pagestyle{plain} 

\setcounter{tocdepth}{2}
{}
\vfill

\ \vspace{-2cm}
\renewcommand{\baselinestretch}{1}\small
\tableofcontents
\renewcommand{\baselinestretch}{1.15}\normalsize

\newcommand{\brho}{\bar{\rho}}
\newcommand{\Svn}{S_{\rm vN}}
\newcommand{\N}{{\cal N}}
\newcommand{\C}{{\cal C}}
\renewcommand{\H}{{\cal H}}

\section{Introduction}

There is an intriguing similarity between tensor networks and emergent geometry in AdS/CFT \cite{Swingle:2009bg}. A quantum state constructed from a random tensor network has an entanglement structure dictated by the geometry of the network, and satisfies a discrete version of the Ryu-Takayanagi formula \cite{Swingle:2009bg,Harlow:2016vwg,Hayden:2016cfa}. There are arguments that a tensor network can be constructed from the bulk theory in principle \cite{Akers:2018fow,Dong:2018seb,VanRaamsdonk:2018zws,Bao:2018pvs,Bao:2019fpq}, and further connections have been developed in, e.g, \cite{Dong:2016eik,Miyaji:2016mxg,Takayanagi:2017knl,Caputa:2017urj,Vasseur:2018gfy,Hayden:2018khn,Alexander:2021jsy,Akers:2021pvd,Cheng:2022ori,Akers:2022zxr,Milekhin:2022zsy,Chen:2022wvy}.

This is closely related to the idea that spacetime can be understood as a quantum error-correcting code  \cite{Verlinde:2012cy,Almheiri:2014lwa,Pastawski:2015qua,Harlow:2016vwg}. A holographic code is a linear map $W$ from the Hilbert space of the bulk low-energy effective field theory to the physical Hilbert space of the dual CFT,
\begin{align}
W: \quad {\cal H}_{\rm EFT} \to {\cal H}_{\rm CFT} \ ,
\end{align}
that preserves some structure of the bulk theory. For example, $W$ must preserve correlation functions of simple operators, such as
\begin{align}
\langle \phi(X_1) \phi(X_2) \rangle_{\rm EFT} = \langle \Phi(X_1) \Phi(X_2) \rangle_{\rm CFT} \ , \quad
\mbox{with} \quad W \phi |0\rangle_{\rm EFT} = \Phi |0\rangle_{\rm CFT} \ ,
\end{align}
where $\phi$ is a local bulk field and $\Phi$  is the corresponding smeared CFT operator. A similar condition applies to higher-point functions. 

One way to satisfy these relations is if $W^{\dagger} W = \id$. A linear map satisfying this condition is said to be an \textit{isometry}, and this defines an isometric code. 
Near the AdS vacuum state, the holographic code is isometric, because bulk operators can be pushed to the boundary with a state-independent smearing kernel \cite{Banks:1998dd,Hamilton:2006az}. 
However, if the $\phi$ operators are hidden behind a horizon, then the code is expected to be non-isometric \cite{Verlinde:2012cy,Marolf:2013dba}. This follows from a simple counting argument: The physical Hilbert space relevant to a black hole has size $e^S$, with $S = \frac{\mbox{area}}{4}$ the Bekenstein-Hawking entropy, but the bulk EFT inside a black hole can have a much larger Hilbert space.  Thus truncating to the relevant parts of the Hilbert spaces we have $\dim({\cal H}_{\rm EFT}) \gg \dim({\cal H}_{\rm CFT})$, and under this condition is is impossible for $W$ to be isometric, because the rank of $W^\dagger W$ is much smaller than its dimension. Random tensor networks exhibit similar behavior:  The code is isometric outside the horizon and non-isometric inside, if a `horizon' is identified as a locally minimal surface in the tensor network \cite{Hayden:2016cfa}.

The correspondence between random tensor networks and AdS/CFT is, for the most part, based on qubit models. In this paper we will study examples where there is a quantitative correspondence to dynamical gravity. We construct (pseudo)random tensor networks, and the corresponding holographic codes, directly from the dual CFT, in the context of the AdS$_3$/CFT$_2$ correspondence and for spherically symmetric states in higher dimensions. These tensor networks can be interpreted as discretizing the bulk radial direction. There are several calculations that support this interpretation. First, the bond dimension of the tensor network agrees with the bulk area at minimal surfaces (but not elsewhere). Second, the resulting codes are isometric outside the horizon, with a transition to non-isometric behavior inside. Third, the replicas built from coarse-grained tensor networks have the same structure as multiboundary wormholes in the bulk.

We consider 3d gravity coupled to point particles, which is conjecturally dual to an ensemble of 2d CFTs with random OPE coefficients \cite{Chandra:2022bqq}. One of our main conclusions is that this model realizes and extends a proposal made in 
\cite{VerlindeBanff,VerlindeInProgress} for how the isometric transition is encoded in the dual CFT (see also \cite{Verlinde:2012cy,Goel:2018ubv,Verlinde:2022xkw}). 
The starting point is that in the high-energy regime, assuming the eigenstate thermalization hypothesis (ETH), probe operators behave like a random map. If a probe operator ${\cal O}$ is dual to a particle outside the horizon, the random map is approximately isometric $M^\dagger M \approx \mathbb{1}$, where $M$ is a finite-dimensional  matrix built by truncating ${\cal O}$ near the semiclassical saddle (and rescaling). But if the probe is inside the horizon, then the map is approximately co-isometric, $MM^\dagger \approx \mathbb{1}$. The transition occurs because particles behind the horizon have negative energy-at-infinity, and the energy controls the effective dimensions and rank of the random map \cite{VerlindeBanff,VerlindeInProgress}. 
As we will review below, this description applies to probes acting on spherically symmetric black holes in an arbitrary number of dimensions.

A more elaborate construction that separates the random nature of black hole microstates from the non-random infrared degrees of freedom is necessary to apply this idea to black holes without spherical symmetry, as we will do here. In spherically symmetric states, it is sufficient to treat the random tensors as acting within the physical CFT Hilbert space (due to Birkhoff's theorem, as discussed below).  In asymmetric states, this does not work; the random tensors must act in an auxiliary Hilbert space. Intuitively, the reason is that light fields in the bulk can carry a leading-order fraction of the total energy, and one must treat the light fields and microscopic degrees of freedom differently in the construction of the tensor network --- the light fields clearly cannot have random matrix elements, so the random tensor bonds correspond only to the microscopic part. It is difficult to build the auxiliary Hilbert space in general, but for a large-$c$ CFT dual to 3d gravity plus massive point particles, the only light field in the bulk is the boundary graviton. In this case the states of the auxiliary Hilbert space are labeled by Virasoro representations. The result is a tensor network that acts within the space of black hole microstates (i.e., heavy primaries), dressed by one final tensor for each boundary component that incorporates the light fields.

The transition in the isometric property at the horizon agrees with general expectations from bulk reconstruction and random tensor networks, which typically deal with small excitations of a given bulk geometry. However, we can go beyond this picture because our starting point is an exact CFT formula for the quantum state, which only reduces to a tensor network near a semiclassical saddle. To illustrate the advantages of the exact formula, we consider the bulk reconstruction of heavy, backreacting operators in 2d CFT. We demonstrate that heavy operators can act isometrically even when they are hidden behind a horizon, and calculate the effective `non-perturbative horizon' defined as the locus where a heavy operator makes the transition to a co-isometric code. This quantifies a sense in which an observer who is allowed to act with heavy operators can easily reconstruct certain operators in a black hole interior.

The results on the isometric transition can also be phrased in terms of identity dominance in the conformal block expansion on the boundary. Correlation functions of point particles in 3d gravity are calculated by Virasoro identity conformal blocks \cite{Hartman:2013mia,Faulkner:2013yia}. That is, gravity calculations are reproduced by terms in the OPE that come from the identity fusion rule,
\begin{align}\label{introOO}
\O^\dagger \O \approx \langle \O^\dagger \O\rangle \id  + \mbox{(Virasoro descendants)} \ .
\end{align}
This resembles the isometry condition for the holographic code, $W^\dagger W \approx \id$. We will show that in our setup they are, in fact, the same: The code is isometric if and only if probe operators satisfy \eqref{introOO} inside arbitrary superpositions of black hole states with a given mass. When the particle dual to ${\cal O}$ is behind the horizon, the Virasoro identity block approximation breaks down due to a counting argument similar to the one above. (A similar breakdown bounds the regime of validity of any bootstrap data extracted by Tauberian methods.) The transition occurs as the dual particle in the bulk is moved across the minimal surface. When the horizon is not spherically symmetric, this translates into a nontrivial property of hyperbolic 2-manifolds, which we prove in section \ref{s:isometric}. 

Even if the operator is behind the horizon, the expectation value of \eqref{introOO} still holds in simple states. This is a CFT realization of bulk reconstruction from non-isometric codes, as anticipated from bulk and information-theoretic arguments in \cite{Akers:2022qdl}, and it is why the Virasoro identity block approximation can be used to calculate correlation functions whether or not the operators are behind a horizon. For large black holes, the approximation is only required to breakdown for very complex states in the black hole interior. However, the breakdown becomes especially severe near a spacetime singularity; see the discussion section for what this means in terms of the Euclidean path integral.

The tensor networks that we construct only discretize the radial direction in the bulk, and only into a finite set of tensor nodes. A very limited  `spatial direction' can be studied in this model by constructing spatial wormholes as 2d CFT tensor networks: Each boundary of the wormholes has a boundary node in the tensor network. In CFT language, this corresponds to creating states by inserting operators on higher genus surfaces. This is not a true discretization of the boundary, but it does allow one to consider the entanglement of boundary subregions in terms of the CFT tensor network, so it is a step in this direction. It is an interesting open question how to construct CFT tensor networks that resolve the spatial directions or allow for a continuum limit of the network.

Section \ref{s:spherical} studies probes of spherically symmetric states in any number of dimensions. Our main new results are described in section \ref{s:rtncft} --- which can be read independently as a technical overview of the paper --- with the supporting gravity calculations on the isometric property given in sections \ref{s:isometric} and \ref{s:btz}. In the discussion section we comment on various open directions, including black hole singularities, finer-grained tensor networks, and corrections to the random tensor approximation required by crossing symmetry.

\subsection{An example}
To illustrate the main ideas, let us consider a pure state in 2d CFT created by the insertion of scalar primary operators inside the unit disk:
\begin{align}
|\Psi\rangle = \O(x)  \O_{i_m}(x_m) \O_{i_{m-1}} (x_{m-1})\cdots \O_{i_2}(x_2) \O_{i_1}(x_1) |0\rangle
\end{align}
Suppose the first $m$ operators are dual to heavy particles, near to but below the black hole threshold.\footnote{Specifically, with weights $h_{i_k} \in (\frac{c}{32}, \frac{c}{24})$ (to avoid complications from multi-twist operators) and in positions such that at the saddlepoint, all internal weights in the comb OPE are above the black hole threshold.} The final operator, $\O$, is special: it has weight $h_\O$ with $1 \ll h_\O \ll \frac{c}{24}$ so it is dual to a light probe particle. 
\definecolor{ctensorcolor}{RGB}{255,50,70}
\definecolor{btensorcolor}{RGB}{0,125,135}
\begin{figure}
\begin{center}
\begin{equation*}
\vcenter{\hbox{
\begin{overpic}[grid=false,width=3.5in]{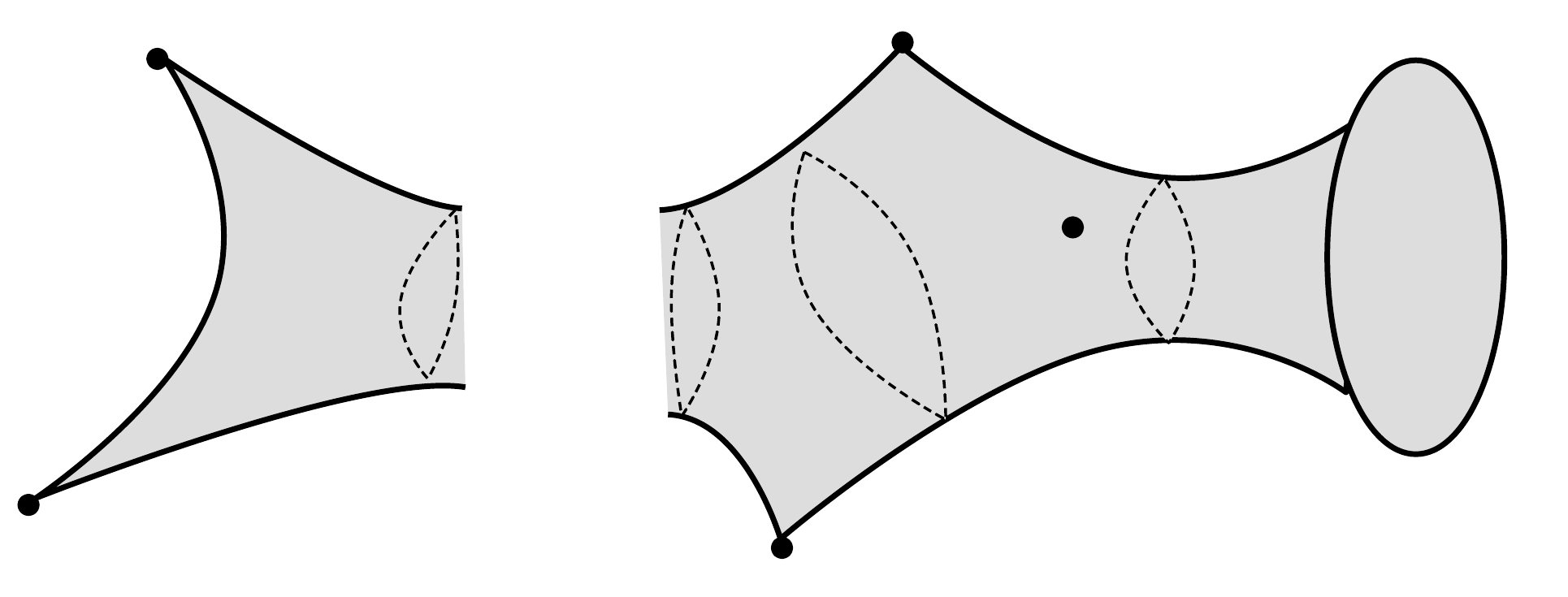}
\put (8,37) {$\mathcal{O}_{i_2}$}
\put (0,0) {$\mathcal{O}_{i_1}$}
\put (48,-2) {$\mathcal{O}_{i_{m-1}}$}
\put (57,38) {$\mathcal{O}_{i_{m}}$}
\put (66,19) {$\mathcal{O}$}
\put (30,20) {$\dots$}
\put (36,20) {$\dots$}
\end{overpic}
}}
\end{equation*}
\begin{tikzpicture}[scale=0.75]
\draw (2,0) -- (2.6,0);
\draw (1.7,0) -- (1.7,1);
\draw (0.7,0) -- (1.7,0);
\draw[black,fill=ctensorcolor] (1.7,0) circle (0.3);
\draw [dashed] (2.8,0) -- (4.3,0);
\draw (4.3,0) -- (5,0);
\draw (5.3,0) -- (5.3,1);
\draw (5.3,0) -- (6.3,0);
\draw[black,fill=ctensorcolor] (5.3,0) circle (0.3);
\draw (6.6,0) -- (6.6,1);
\draw (5.6,0) -- (6.6,0);
\draw (6.6,0) -- (7.6,0);
\draw[black,fill=ctensorcolor] (6.6,0) circle (0.3);
\draw (7.9,0) -- (7.9,1);
\draw (7.9,0) -- (8.9,0);
\draw[black,fill=ctensorcolor] (7.9,0) circle (0.3);
\begin{scope}
\clip (1+7.9,-1) rectangle (1.5+7.9,1);
\draw[black,fill=btensorcolor] (1.5+7.9,0) circle (0.5);
\end{scope}
\draw (1.5+7.9, -.05) -- (2.5+7.9,-.05);
\draw (1.5+7.9, .05) -- (2.5+7.9,.05);
\draw[thick] (1.5+7.9,-0.5) -- (1.5+7.9,0.5);
\node[above] at (5.3,1) {$i_{m-1}$};
\node[above] at (6.6,1) {$i_m$};
\node[above] at (7.9,1) {$\mathcal{O}$};
\node[above] at (1.7,1) {$i_2$};
\node[left] at (0.7,0) {$i_1$};
\node[below] at (2.3,0) {$E_1$};
\node[below] at (6.2,0) {$E_{m-2}$};
\node[below] at (7.4,0) {$E_{m-1}$};
\node[below] at (8.6,0) {$E_m$};
\end{tikzpicture}
\end{center}
\caption{\small\label{fig:introexample}
The spatial geometry of a pure-state black hole in AdS$_3$ and the corresponding pseudorandom tensor network. The tensors, which are defined in terms of OPE coefficients (circles) and Virasoro OPE blocks (semicircles), discretize the radial direction in the bulk. 
}
\end{figure}
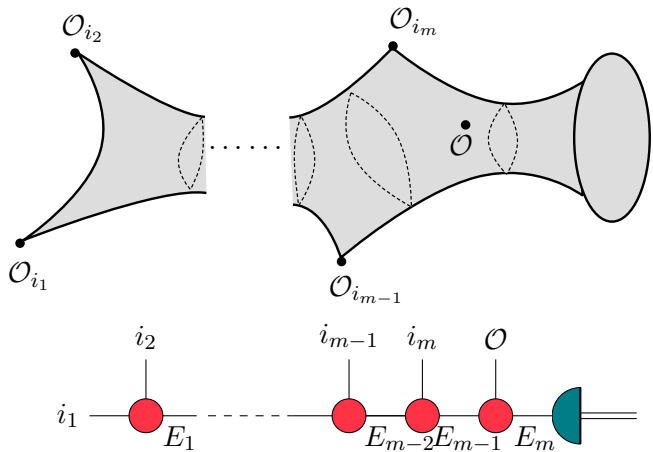

Under these conditions, the state created on the unit circle in radial quantization is dual to a black hole. The $t=0$ spatial geometry and the corresponding tensor network are shown in figure \ref{fig:introexample}. The tensor network is not the exact state, but a truncated version with the sum over internal weights in the OPE limited to states near the semiclassical saddle in $\langle \Psi| \Psi\rangle$; the tensor network state $|\Psi\rangle_*$ is dual to a fixed-area state in the bulk \cite{Akers:2018fow,Dong:2018seb}. There is a precise formula for each node in the network (see section \ref{s:rtncft}), up to an undetermined psuedorandom tensor with zero mean and unit variance, and calculations done with the tensor network match quantitatively to the bulk. The internal bonds are labeled by primaries, and the red tensors are finite-dimensional, with entries proportional to the primary OPE coefficients $c_{pqr}$. The final tensor on the right is a Virasoro OPE block that maps $\H_{\rm primaries} \to \H_{\rm CFT}$ by dressing the primary state with descendants. 

Each extremal surface in the bulk has a corresponding internal line in the network, with bond dimension $e^{S(E_i^*)}$, where $S(E_i^*)$ is the Cardy entropy at the primary weight that appears in the OPE at the saddlepoint (and $E=2h$). For each of these bonds, $S(E_i^*) = \frac{1}{4}\mbox{Area}$. Due to the light probe there is also an extra internal line in the network whose entropy does not correspond to any bulk area. 

In this context, the holographic code $W$ maps the labels on the  operators, $\{i_1,x_1;  i_2, x_2; \dots \}$, into the Hilbert space of the dual CFT. Since this map passes through the node corresponding to the probe insertion $\O$, the code can only be isometric if each tensor, viewed as a linear map from left to right in the figure, is isometric. In particular for $W$ to be isometric, the tensor dual to the probe operator $\O$ must act isometrically. Since this is a random map, it is approximately isometric or co-isometric depending on whether the saddlepoint entropies increase or decrease at this node. We will match this behavior to the bulk by showing that $\O$ acts isometrically when the dual probe particle is outside the extremal surface, and co-isometrically when it is inside. This black hole is not spherically symmetric; the extremal surface is a geodesic in the 2d hyperbolic metric on the unit disk with conical defects at the operator insertions, and the agreement holds everywhere along this curve.

In terms of the Virasoro identity block approximation, the statement is as follows: If the primary weights at the saddle satisfy $E_m < E_{m-1}$ --- implying that the code is non-isometric --- then there exist superpositions of the form $|a\rangle = \sum_{\{i_k\}} a_{i_1\cdots i_n}\O_{i_n}(x_n) \cdots \O_{i_1}(x_1)|0\rangle$, which have the same bulk geometry as $|\Psi\rangle$ outside the outermost horizon, such that the probe correlation function $\langle a| \O(x)^\dagger \O(x) |a\rangle$ differs at leading order from the identity approximation. In fact, there must exist such states that are annihilated by $\O(x)$ to leading order, because the operator $\O^\dagger \O$ (viewed as a finite-dimensional matrix acting on states near the semiclassical saddle) has rank less than its dimension. If we assume the CFT has a large number of flavors, so the heavy operators are labeled by $i=1\dots  N_f $ with $N_f> e^S$, then these $|a\rangle$ states can (in principle) be found by fixing the operator locations and taking a superposition over flavors (similar to \cite{Penington:2019kki}). Otherwise, we can build superpositions with a large number of heavy operators inserted far behind the horizon (i.e., near the origin in CFT language). The exact details of the states that violate the identity approximation cannot be found without knowing the precise OPE coefficients in the CFT, but the counting argument shows that they must exist.

\section{Spherically symmetric states}\label{s:spherical}
In this section we consider geodesic probes of spherically symmetric black holes in AdS$_{d+1}$/CFT$_d$, for any $d \geq 2$. This is largely a review of results described in \cite{VerlindeBanff,Verlinde:2022xkw,VerlindeInProgress}, rephrased in the language of random tensor networks and for pure states rather than eternal black holes (which can be treated similarly). For concreteness we will consider pure state black holes created by a thin shell of matter, but the discussion also applies to other types of matter, such as end-of-the-world branes.

\subsection{The probe OPE as a random tensor network}

Let $V$ be a CFT operator that creates a spherically symmetric thin shell of matter, and $|S\rangle = V |0\rangle$ the non-normalizable CFT state at the shell insertion. These states and their holographic duals are studied in detail in \cite{Anous:2016kss,Chandra:2022fwi,Bah:2022uyz}. Evolving in Euclidean time prepares a normalizable state, $ e^{-\tau_0 H} |S\rangle$.
We assume the mass of the shell is large enough so that the shell is behind the horizon at $t=0$.
Now let us act on this state with additional probe operators,
\begin{align}\label{probestate}
|\Psi\rangle &= {\cal O}^{i_m}(-\tau_m)\cdots {\cal O}^{i_2}(-\tau_2) {\cal O}^{i_1}(-\tau_1) e^{-\tau_0 H} |S\rangle \ ,
\end{align}
with the operators ordered in Euclidean time,
\begin{align}
\tau_0 > \tau_1 > \tau_2 > \cdots > \tau_m > 0 \ .
\end{align}
The ${\cal O}^i$ are scalar primary operators, with $i$ a flavor index, and scaling dimensions satisfying $1 \ll \Delta_i \ll N^2$. These are dual to massive probe particles in the bulk, which travel on geodesics. For example, the Euclidean geometry dual to $\langle \Psi | \Psi\rangle$ for $m=5$, with two probes behind the horizon and three outside, is
\begin{align}\label{manyprobes}
\vcenter{\hbox{
\begin{overpic}[grid=false,width=1.5in]{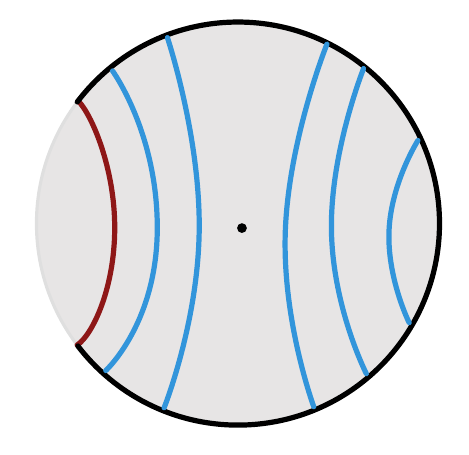}
\put (8,78) {$\tau_0$}
\put (16,85) {$\tau_1$}
\put (29,93) {$\tau_2$}
\put (65,92) {$\tau_3$}
\put (78,84) {$\tau_4$}
\put (90,70) {$\tau_5$}
\end{overpic}
}}
\end{align}
where the red curve is the spherically symmetric thin shell, and the blue curves are the geodesics of the massive probe particles. The figure shows the radial direction and Euclidean time. Only the black hole portion is drawn; this is a pure state, so it is glued to vacuum global AdS at the shell. For more details, including the solution to the shell EOM and the expansion of $|S\rangle$ in CFT eigenstates, see \cite{Chandra:2022fwi}.

The spatial geometry of the $t=0$ slice has a spherical shell behind a minimal surface. Schematically, it looks like:
\begin{align}\label{manyprobesslice}
\vcenter{\hbox{
\includegraphics{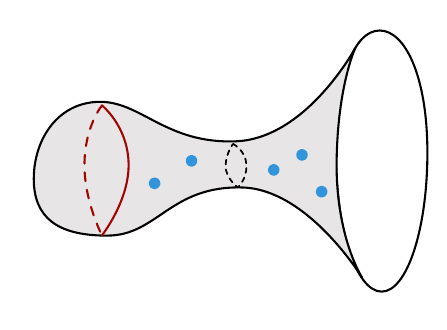}
}}
\end{align}
Here we show the radial direction and the transverse directions, $S^{d-1}$. We will recast the CFT state as a random tensor network with a network geometry that resembles \eqref{manyprobesslice}.  The tensors discretize the radial direction, with the rank of the tensor bonds related to the transverse area. Locality in the transverse directions does not play any role, so to simplify the discussion we restrict to the zero-momentum sector by integrating the probes over the spatial sphere, choosing ${\cal O}^i(\tau_i) = \int d^{d-1}x{\cal O}^i(\tau_i, \vec{x}_i)$.  (In 2d CFT we will consider local operators below.)

By inserting complete sets of energy eigenstates, the exact CFT state \eqref{probestate} can be expressed diagrammatically as
\begin{align}\label{exactstatediagram}
\vspace{1cm}|\Psi\rangle &= \vcenter{\hbox{
	\begin{tikzpicture}[scale=1]
	\draw[thick] (0.5,0) circle (0.5);
	\draw[thick] (1,0) -- (2,0);
	\draw[thick] (2,-0.5) -- (2,0.5);
	\draw[thick] (2.5,0.5) -- (2.5,1);
	\draw[thick] (2,0.5) -- (3,0.5);
	\draw[thick] (3,0.5) -- (3,-0.5);
	\draw[thick] (2,-0.5) -- (3,-0.5);
	\draw[thick] (3,0) -- (3.5,0);
	\draw[dashed] (3.5,0) -- (4.5,0);
	\draw[thick] (4.5,0) -- (5,0);
	\draw[thick] (5,-0.5) -- (5,0.5);
	\draw[thick] (5,0.5) -- (6,0.5);
	\draw[thick] (5.5,0.5) -- (5.5,1);
	\draw[thick] (6,-0.5) -- (6,0.5);
	\draw[thick] (5,-0.5) -- (6,-0.5);
	\draw[thick] (6,0) -- (7,0);
	\draw[thick] (7,-0.5) -- (7,0.5);
	\draw[thick] (7.5,0.5) -- (7.5,1);
	\draw[thick] (7,0.5) -- (8,0.5);
	\draw[thick] (8,-0.5) -- (8,0.5);
	\draw[thick] (7,-0.5) -- (8,-0.5);
	\draw[thick] (8,0) -- (9,0);
	\node[scale=0.85] at (0.5,0) {$-\tau_0$};
	\node[scale=0.85] at (2.5,0) {$-\tau_1$};
	\node[scale=0.85] at (5.5,0) {$-\tau_{m-1}$};
	\node[scale=0.85] at (7.5,0) {$-\tau_{m}$};
	\node[above] at (2.5,1) {$i_1$};
	\node[above] at (5.5,1) {$i_{m-1}$};
	\node[above] at (7.5,1) {$i_m$};
	\end{tikzpicture}
	}}
\end{align}
where we have defined the tensors
\begin{align}
\vcenter{\hbox{
	\begin{tikzpicture}[scale=1]
	\draw[thick] (0.5,0) circle (0.5);
	\draw[thick] (1,0) -- (2,0);
         \node[right] at (2,0) {$n$};
       \node[scale=0.85] at (0.5,0) {$-\tau_0$};
        \end{tikzpicture}
}}
&= e^{-\tau_0 E_n} \langle n | S \rangle \\
\vcenter{\hbox{
\begin{tikzpicture}[scale=1]
	\draw[thick] (1,0) -- (2,0);
	\draw[thick] (2,-0.5) -- (2,0.5);
	\draw[thick] (2.5,0.5) -- (2.5,1);
	\draw[thick] (2,0.5) -- (3,0.5);
	\draw[thick] (3,0.5) -- (3,-0.5);
	\draw[thick] (2,-0.5) -- (3,-0.5);
	\draw[thick] (3,0) -- (4,0);
        \node[above] at (2.5,1) {$i$};
        \node[scale=.85] at (2.5,0) {$\tau$};
        \node[left] at (1,0) {$m$};
        \node[right] at (4,0) {$n$};
 \end{tikzpicture}
}}
 &= \langle n | \O^i(\tau)|m\rangle \ .
\end{align}
The tensors $\O^i_{mn}$ are infinite dimensional in the lower indices, indexed by energy eigenstates $|m\rangle$, $|n\rangle$. Connected lines between tensors are contracted indices, and the free line at the right end of \eqref{exactstatediagram} corresponds to an uncontracted index in the physical Hilbert space.

In a theory satisfying the eigenstate thermalization hypothesis (ETH), the matrix elements of a light probe between two high-energy eigenstates can be approximated by\footnote{The ETH in QFT must also account for momentum conservation. In \eqref{oeth} we implicitly assume that $|m\rangle$ and $|n\rangle$ have equal momentum, since ${\cal O}$ is averaged over the sphere.}
\begin{align}\label{oeth}
\langle m|{\cal O}^i|n\rangle &= 
\langle{\cal O}^i\rangle_{\beta(E_m)}\delta_{mn} + C^i(E_m, E_n)R^i_{mn}
\end{align}
where $R^i$ is a random matrix with zero mean and unit variance, and $C^i(E,E')$ is a smooth function of energies determined by matching this ansatz to the thermal 2-point function.  We assume the thermal 1-point vanishes, so we can drop the first term in \eqref{oeth}.

Applying the ETH to \eqref{exactstatediagram}, the state $|\Psi\rangle$ becomes a weighted random tensor network with weights determined by the thermal 2-point functions. The tensors are infinite dimensional, but calculations are often dominated by a semiclassical saddlepoint, and then the tensors effectively become finite dimensional. Suppose the sum over energies in the spectral decomposition is dominated by saddlepoint energies, $E^*_k$. Then we can truncate the sums to a microcanonical window of $e^{S_k}$ states around the saddle, with $S_k = S(E_k^*)$ the saddlepoint entropy. The tensor for $\O(-\tau_k)$ becomes a rectangular matrix of dimensions $e^{S_{k-1}} \times e^{S_{k}}$. This effective dimension only makes sense in the vicinity of a given saddlepoint.

Upon doing this truncation, the resulting finite-dimensional tensor network resembles the bulk spatial geometry \eqref{manyprobesslice}, with the rank of the tensors playing the role of the transverse area. Consider the norm,
\begin{align}
\langle \Psi | \Psi\rangle  = 
\vcenter{\hbox{
	\begin{tikzpicture}[scale=1]
	\draw[thick] (0.5,0) circle (0.5);
	\draw[thick] (1,0) -- (2,0);
	\draw[thick] (2,-0.5) -- (2,0.5);
	\draw[thick] (2.5,0.5) -- (2.5,1);
	\draw[thick] (2,0.5) -- (3,0.5);
	\draw[thick] (3,0.5) -- (3,-0.5);
	\draw[thick] (2,-0.5) -- (3,-0.5);
	\draw[thick] (3,0) -- (3.5,0);
	\draw[dashed] (3.5,0) -- (4.5,0);
	\draw[thick] (4.5,0) -- (5,0);
	\draw[thick] (5,-0.5) -- (5,0.5);
	\draw[thick] (5,0.5) -- (6,0.5);
	\draw[thick] (5.5,0.5) -- (5.5,1);
	\draw[thick] (6,-0.5) -- (6,0.5);
	\draw[thick] (5,-0.5) -- (6,-0.5);
	\draw[thick] (6,0) -- (7,0);
        \draw[thick] (7,-0.5) -- (7,0.5);
	\draw[thick] (7,0.5) -- (8,0.5);
	\draw[thick] (7.5,0.5) -- (7.5,1);
	\draw[thick] (8,-0.5) -- (8,0.5);
	\draw[thick] (7,-0.5) -- (8,-0.5);
        \draw[thick] (8,0) -- (8.5,0);
        \draw[dashed] (8.5,0) -- (9.5,0);
        \draw[thick] (9.5,0) -- (10,0);
         \draw[thick] (10,-0.5) -- (10,0.5);
	\draw[thick] (10,0.5) -- (11,0.5);
	\draw[thick] (10.5,0.5) -- (10.5,1);
	\draw[thick] (11,-0.5) -- (11,0.5);
	\draw[thick] (10,-0.5) -- (11,-0.5);
        \draw[thick] (11,0) -- (12,0);
         \draw[thick] (12.5,0) circle (0.5);
	\node[scale=0.8] at (0.5,0) {$-\tau_0$};
	\node[scale=0.8] at (2.5,0) {$-\tau_1$};
	\node[scale=0.8] at (5.5,0) {$-\tau_{m}$};
	\node[above,scale=0.85] at (2.5,1) {$i_1$};
	\node[above,scale=0.85] at (5.5,1) {$i_{m}$};
        \node[above,scale=0.85] at (7.5,1) {$i_m$};
       \node[above,scale=0.85] at (10.5,1) {$i_1$};
       \node[scale=0.8] at (12.5,0) {$\tau_0$};
        \node[scale=0.8] at (10.5,0) {$\tau_1$};
        \node[scale=0.8] at (7.5,0) {$\tau_m$};
	\end{tikzpicture}
	}}
\end{align}

The saddlepoint in the sum over contracted indices, with saddlepoint energies $E_k^*$ for $k=0,\dots,m$ (corresponding to the internal legs from left to right in the diagram), is dual to the Euclidean spacetime \eqref{manyprobes}. The geometry of the network matches the geometry of the bulk spatial slice, in the sense that the saddlepoint entropies increase/decrease along the tensor network with the same pattern that the transverse area increases/decreases along the radial direction in \eqref{manyprobesslice}.
At the minimal surface,  the transverse area matches the tensor rank on that leg, i.e. $\log \dim {\cal H}_{eff} = S(E^*)=\frac{\mbox{area}}{4}$.

From this correspondence we can understand the isometric property of the holographic code by following \cite{VerlindeBanff,Verlinde:2022xkw,VerlindeInProgress}. If ${\cal O}_i$ is behind the apparent horizon, then it decreases the saddlepoint energy, and if it is outside the apparent horizon, it increases the saddlepoint energy. The entropies also satisfy this hierarchy,
\begin{align}\label{sphericalhierarchy}
{\cal O}^i \mbox{\ inside\ } &\quad \Rightarrow \quad E_i^* < E_{i-1}^*, \quad S_i < S_{i-1} \\
{\cal O}^i \mbox{\ outside\ } &\quad \Rightarrow \quad E_i^* > E_{i-1}^*, \quad S_i > S_{i-1}\notag
\end{align}
These inequalities are derived from a straightforward bulk argument reviewed in section \ref{ss:sphericalderivation} below.

The properties of a random map depend crucially on whether it maps a smaller space to a larger space, or vice-versa.
Truncating to a microcanonical window near the saddlepoint and assuming ETH, each probe operator ${\cal O}$ acts like a finite-dimensional random map,
\begin{align}
\O(-\tau_k) :  \quad {\cal H}(E_{k-1}^*) \to {\cal H}(E_k^*)
\end{align}
where ${\cal H}(E)$ is the Hilbert space consisting of $e^{S(E)}$ CFT states around energy $E$. For an operator outside the horizon, this is a random map from smaller space to a larger one, so it acts invertibly, $\O^\dagger \O \propto \id$. By contrast, for a particle behind the horizon, ${\cal O}$ maps a larger space to a smaller one, so it cannot be invertible. That is, the rank of $\left(\O(-\tau_k)\right)^\dagger \O(-\tau_k)$ (with both operators truncated to finite-dimensional matrices around the saddle) is bounded above by $e^{S_k}$, which is less than its dimension if $S_{k} < S_{k-1}$. 

For illustration, consider the particular state illustrated in \eqref{manyprobes}, which has five probe particles --- two behind the horizon, and three outside the horizon. The state is
\begin{align}
|\Psi\rangle = \vcenter{\hbox{
	\begin{tikzpicture}[scale=1]
	\draw[thick] (0.5,0) circle (0.5);
	\draw[thick] (1,0) -- (2,0);
	\draw[thick] (2,-0.5) -- (2,0.5);
	\draw[thick] (2.5,0.5) -- (2.5,1);
	\draw[thick] (2,0.5) -- (3,0.5);
	\draw[thick] (3,0.5) -- (3,-0.5);
	\draw[thick] (2,-0.5) -- (3,-0.5);
	\draw[thick] (3,0) -- (4,0);
        \draw[thick] (4,-0.5) -- (4,0.5);
	\draw[thick] (4.5,0.5) -- (4.5,1);
	\draw[thick] (4,0.5) -- (5,0.5);
	\draw[thick] (5,0.5) -- (5,-0.5);
	\draw[thick] (4,-0.5) -- (5,-0.5);
	\draw[thick] (5,0) -- (6,0);
	 \draw[thick] (6,-0.5) -- (6,0.5);
	\draw[thick] (6.5,0.5) -- (6.5,1);
	\draw[thick] (6,0.5) -- (7,0.5);
	\draw[thick] (7,0.5) -- (7,-0.5);
	\draw[thick] (6,-0.5) -- (7,-0.5);
	\draw[thick] (7,0) -- (8,0);
         \draw[thick] (8,-0.5) -- (8,0.5);
	\draw[thick] (8.5,0.5) -- (8.5,1);
	\draw[thick] (8,0.5) -- (9,0.5);
	\draw[thick] (9,0.5) -- (9,-0.5);
	\draw[thick] (8,-0.5) -- (9,-0.5);
	\draw[thick] (9,0) -- (10,0);
         \draw[thick] (10,-0.5) -- (10,0.5);
	\draw[thick] (10.5,0.5) -- (10.5,1);
	\draw[thick] (10,0.5) -- (11,0.5);
	\draw[thick] (11,0.5) -- (11,-0.5);
	\draw[thick] (10,-0.5) -- (11,-0.5);
	\draw[thick] (11,0) -- (12,0);
	\node[scale=0.85] at (0.5,0) {$-\tau_0$};
	\node[scale=0.85] at (2.5,0) {$-\tau_1$};
	\node[scale=0.85] at (4.5,0) {$-\tau_2$};
	\node[scale=0.85] at (6.5,0) {$-\tau_3$};
        \node[scale=0.85] at (8.5,0) {$-\tau_4$};
        \node[scale=0.85] at (10.5,0) {$-\tau_5$};
	\node[above] at (2.5,1) {$i_1$};
	\node[above] at (4.5,1) {$i_2$};
	\node[above] at (6.5,1) {$i_3$};
        \node[above] at (8.5,1) {$i_4$};
        \node[above] at (10.5,1) {$i_5$};
        \node[below] at (1.5,0) {0};
        \node[below] at (3.5,0) {1};
        \node[below] at (5.5,0) {2};
         \node[below] at (7.5,0) {3};
          \node[below] at (9.5,0) {4};
        \node[below] at (11.5,0) {5};
	\end{tikzpicture}
	}}
\end{align}
Applying \eqref{sphericalhierarchy}, we see that the saddlepoint entropies in $\langle \Psi | \Psi\rangle$ satisfy 
\begin{align}
S_0 > S_1 > S_2 , \quad S_2 < S_3 < S_4 < S_5 \ .
\end{align}
This hierarchy of entropies matches the geometry of the spatial slice drawn in \eqref{manyprobesslice}, and the minimal entropy, $S_2$, matches the area of the minimal surface in \eqref{manyprobes}. The other entropies, away from the minimal one, do not satisfy $S = \frac{\mbox{area}}{4}$ for the corresponding bulk region, but they increase/decrease in the same pattern as the transverse area. This is the sense in which the spatial geometry is discretized by a random tensor network.

\subsubsection*{The role of Birkhoff's theorem}
It was essential to this entire discussion that we assumed spherical symmetry. At a technical level, it enters in the derivation of the energy hierarchy below. But this is more than just a technical simplification. Birkhoff's theorem states that the bulk geometry of a spherically symmetric state is locally identical to an eternal black hole. In CFT language, this means that there is no need to keep track of anything besides the energy, and consequently, we can label internal lines in the tensor network by physical CFT states, as was done above. As long as we assign the correct energies to each leg, they will also have the correct entropy, so the tensor ranks along the tensor network match the areas along the bulk radial direction. Additional spherical shell operators can be incorporated without any major differences.

For states without spherical symmetry, the story is much more subtle. General arguments indicate that a probe operator should transition between isometric/non-isometric at the minimal surface, but without spherical symmetry, the minimal surface is \textit{not} where the energy hierarchy inverts. As we will see in 3d gravity, the reason for the mismatch is that generally the Hilbert space assigned to the tensor legs is not the physical CFT Hilbert space --- it is an auxiliary Hilbert space with a reduced number of states. The isometric property must be understood in terms of counting states in the auxiliary Hilbert space. We will see how this works explicitly in 2d CFT and find that when the state counting is done in the auxiliary Hilbert space, the isometric property matches precisely with the location of the minimal surface.

\subsection{Derivation of the energy hierarchy with spherical symmetry}\label{ss:sphericalderivation}

It remains to establish the inequalities in \eqref{sphericalhierarchy} for the saddlepoint energies in spherically symmetric states. Similar calculations were done in \cite{VerlindeBanff,Verlinde:2022xkw,VerlindeInProgress}.   The smooth function $C^i(E,E')$ in the matrix elements of the CFT operators are determined by matching to the bulk, so this calculation can be done on the gravity side --- the CFT is guaranteed to agree.  

Consider the state \eqref{probestate} with $m$ probe operators. Choose a bulk radial slice $\tau = \tau_c$ in the Euclidean spacetime, with $\tau_{i+1} < \tau_c < \tau_{i}$, $i \geq 1$ (and define $\tau_{m+1} = 0$). The ADM energy on this slice, $E(\tau_c)$, is equal to the saddlepoint energy in the CFT spectral expansion for the intermediate state running between the operators $\O_{i}$ and $\O_{i+1}$. Now we will compare this to the energy $E(\tau_c')$ for a slice with $\tau_{i} < \tau_{c}' < \tau_{i-1}$. 

\subsubsection*{$\O_i$ outside}
If the particle dual to $\O_i$ is outside the horizon, then the situation looks like this:
\begin{align}
\vcenter{\hbox{
\begin{overpic}[grid=false, width=1.3in]{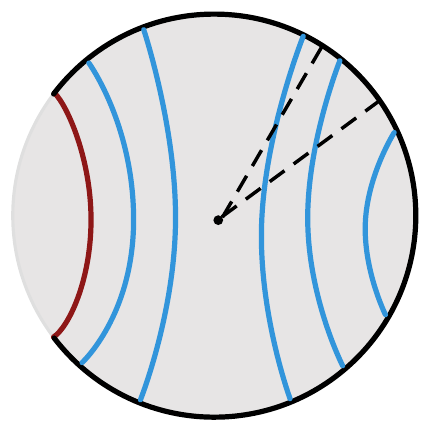}
\put (90,78) {$\tau_c$}
\put (74,93) {$\tau_c'$}
\put (60,-3) {{\footnotesize $i-1$}}
\put (79,7) {{\footnotesize $i$}}
\put (90,19) {{\footnotesize $i+1$}}
\end{overpic}
}}
\end{align}
To first order in the backreaction, the ADM energy is that of the black hole plus an $O(m)$ term from each particle on the slice. The slice $\tau = \tau_c$ contains an extra particle compared to $\tau = \tau_c'$, so it has higher energy: $E(\tau_c) > E(\tau_c')$. Translating this into the CFT saddlepoint energies we have shown
\begin{align}
E_i^* > E_{i-1}^* \ .
\end{align}
This can also be phrased in terms of energy flux into the boundary. The boundary stress tensor $T_{ij}$ satisfies the conservation law \cite{Brown:1992br}
\begin{align}\label{bycons}
\del_i T^{ij} = -n_\mu T_{\rm bulk}^{\mu j} \ ,
\end{align}
where $\mu$ is a bulk index, $i$ is a boundary index, $n$ is the unit normal to the boundary, and $T^{\mu\nu}_{\rm bulk}$ is the matter stress tensor in the bulk. As we evolve from the $\tau_c'$ slice to the $\tau_c$ slice, a positive-energy particle enters through the boundary, providing a positive flux in \eqref{bycons} and thereby increasing the ADM energy.

\subsubsection*{$\O_i$ inside}
If the particle dual to $\O_i$ is inside the horizon, then instead it looks like this:
\begin{align}
\vcenter{\hbox{
\begin{overpic}[grid=false, width=1.3in]{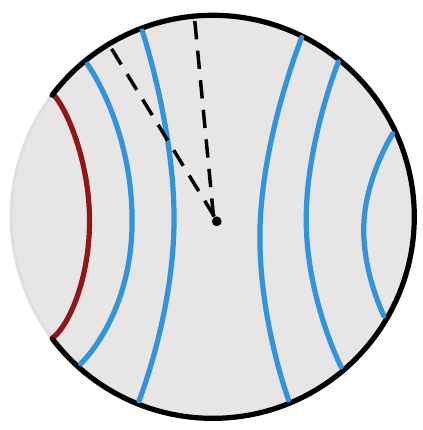}
\put (41,100) {$\tau_c$}
\put (17,94) {$\tau_c'$}
\put (2,6) {{\footnotesize $i-1$}}
\put (27,-3) {{\footnotesize $i$}}
\put (60,-3) {{\footnotesize $i+1$}}
\end{overpic}
}}
\end{align}
Now it is the slice $\tau = \tau_c'$ that has an extra particle. Therefore, by the same argument, 
\begin{align}
E_i^* < E_{i-1}^* \ .
\end{align}
In this case the particle exits through the boundary, so the flux is negative.

\section{Random tensor networks from 2d CFT}\label{s:rtncft}

In \cite{Chandra:2022bqq} it was argued that 3d gravity coupled to massive particles is dual to an ensemble of large-$c$ CFTs with random OPE coefficients. 
In related work, we showed that multi-boundary wormholes (in any dimension) can be interpreted as replica partition functions for coarse-grained states, in a fixed theory \cite{Chandra:2022fwi}. Here we will show that this same model in 2d CFT also leads to a correspondence between Virasoro OPE blocks, random tensor networks, and bulk spatial geometries. In this section we describe the CFT construction. We will use some 3d gravity results as input, postponing the details of the gravity calculations to section \ref{s:isometric} below.

We assume the CFT spectrum consists of a small number of single-trace primary operators ${\cal O}^i$ below the black hole threshold, their multi-trace composites, and a Cardy spectrum of black hole microstates above the threshold, with $h,\bh > \frac{c}{24}$.

\subsection{Virasoro OPE blocks}
\newcommand{\B}{\mathcal{B}}

On the Euclidean cylinder with coordinates $(\tau,\phi)$, consider the state
\begin{align}\label{psi2d}
|\Psi\rangle = \O^{i_m}(-\tau_m,\phi_m)  \cdots  \O^{i_2}(-\tau_2, \phi_2) \O^{i_1}(-\tau_1,\phi_1)|0\rangle
\end{align}
where the $\O^i$ are scalar primaries and $\tau_1 > \tau_2 > \cdots > \tau_m$. 
The spectrum decomposes into Virasoro representations,
\begin{align}
\mathbb{1} = \sum_n |n\rangle\langle n| = \sum_{p} \P_p \ , 
\end{align}
where $n$ runs over all states, $p$ runs over primaries, and $\P_p$ is the projector onto the representation with lowest weight $p$.\footnote{I.e., $\P_p := \sum_{M,N} L_M \bar{L}_N |p\rangle \langle p| L_M^{\dagger}\bar{L}_N^\dagger$ with $|p\rangle$ the primary state and $L_M$ and $\bar{L}_N$ the orthonormalized chiral and anti-chiral raising operators built from Virasoro modes.} Inserting this into \eqref{psi2d} gives an expansion in Virasoro OPE blocks: 
 \begin{align}\label{Bexp}
|\Psi\rangle = \sum_{p_1,\dots, p_{m-1}} c_{i_1 i_2 p_1} \cdots c_{{i_m}p_{m-2} p_{m-1}} 
\left| \B\left[ \vcenter{\hbox{\vspace{0.12in}
	\begin{tikzpicture}[scale=.75]
        \draw[thick,->] (3/2,0) -- (5/2,0);
        \node[below,scale=0.75] at (2.2,0) {$h_{p_{m-1}}$};
        \draw[thick] (3/2,0) -- (3/2,0.8);
        \node[above,scale=0.75] at (3/2,0.8) {$h_{i_m}$};
        \draw[thick] (1/2,0) -- (3/2,0); 
	\draw[dashed] (-1/2,0) -- (1/2,0);
	\node[below,scale=0.75] at (1,0) {$h_{p_{m-2}}$};
	\draw[thick] (-1/2,0) -- (-3/2,0);
	\node[below,scale=0.75] at (-1,0) {$h_{p_1}$};
	\draw[thick] (-1/2,0) -- (-1/2,1*0.8);
	\node[above,scale=0.75] at (-1/2,0.8) {$h_{i_3}$};
	\draw[thick] (-3/2,0) -- (-3/2,0.8);
	\node[above,scale=0.75] at (-3/2,0.8) {$h_{i_2}$};
        \draw[thick] (-3/2,0) -- (-5/2,0);
        \node[left,scale=0.75] at (-5/2,0) {$h_{i_1}$};
	\end{tikzpicture}
	}} \right] \right|^2|p_{m-1}\rangle 
\end{align}
The notation ${\cal B}[\dots]$ represents a chiral Virasoro OPE block, which is defined by this equation; ${\cal B}$ is the contribution to the OPE with the given primary labels, with OPE coefficients stripped off. 
It is an operator that acts within a fixed representation by sums of products of the Virasoro raising operators $L_{-n}$ for $n\geq 1$, is completely determined by the Virasoro algebra, and depends holomorphically on the weights $\{ h_{i_k} \}, \{h_{p_k}\}$ and positions $\{z_k\}$. In the diagram, the arrow shows which leg is acting as an operator, so in this case the operator acts within the representation with lowest weight $(h_{p_{m-1}}, \bh_{p_{m-1}})$.\footnote{For a general discussion of OPE blocks see \cite{Czech:2016xec} where this terminology was first introduced. See also \cite{Fitzpatrick:2016mtp} for Virasoro OPE blocks and their relation to 3d gravity. Our convention is to include all of the position dependence in the OPE blocks and similarly for conformal blocks.}

The norm $\langle \Psi| \Psi\rangle$ is a $2m$-point correlation function. It can be expanded in Virasoro conformal blocks, 
\begin{align}
\langle \Psi | \Psi\rangle &= \sum_{p_1, \dots, p_{2m-3}} 
 c_{i_1 i_2 p_1}\cdots c^*_{i_{2m}i_{2m-1}p_{2m-3}} \left |
 \vcenter{\hbox{\vspace{0.12in}
	\begin{tikzpicture}[scale=.75]
	\draw[thick] (3/2,0) -- (5/2,0);
        \node[right,scale=0.75] at (5/2,0) {$i_{2m}$};
        \draw[thick] (3/2,0) -- (3/2,0.8);
        \node[above,scale=0.75] at (3/2,0.8) {$i_{2m-1}$};
        \draw[thick] (1/2,0) -- (3/2,0); 
	\draw[dashed] (-1/2,0) -- (1/2,0);
	\node[below,scale=0.75] at (1,0) {$p_{2m-3}$};
	\draw[thick] (-1/2,0) -- (-3/2,0);
	\node[below,scale=0.75] at (-1,0) {$p_1$};
	\draw[thick] (-3/2,0) -- (-3/2,0.8);
	\node[above,scale=0.75] at (-3/2,0.8) {$i_2$};
        \draw[thick] (-3/2,0) -- (-5/2,0);
        \node[left,scale=0.75] at (-5/2,0) {$i_1$};
	\end{tikzpicture}
	}}\right |^2
 \end{align}
This comb diagram represents the usual chiral Virasoro conformal block ${\cal F}$. It is related to the OPE block by ${\cal F} = \langle p| {\cal B}^\dagger {\cal B} |p \rangle$; for example for $4$-point functions the relation is
\begin{align}\label{pFFp}
  \vcenter{\hbox{\vspace{0.12in}
	\begin{tikzpicture}[scale=.75]
        \draw[thick] (-1/2,0) -- (-1/2,0.8);
        \node[above, scale=0.75] at (-1/2,0.8) {$i_3$};
        \draw[thick] (-1/2,0) -- (1/2,0);
        \node[right, scale=0.75] at (1/2,0) {$i_4$};
	\draw[thick] (-1/2,0) -- (-3/2,0);
	\node[below,scale=0.75] at (-1,0) {$p$};
	\draw[thick] (-3/2,0) -- (-3/2,0.8);
	\node[above,scale=0.75] at (-3/2,0.8) {$i_2$};
        \draw[thick] (-3/2,0) -- (-5/2,0);
        \node[left,scale=0.75] at (-5/2,0) {$i_1$};
	\end{tikzpicture}
	}} = \langle p | {\cal B}\left[  \vcenter{\hbox{\vspace{0.12in}
	\begin{tikzpicture}[scale=.75]
	\draw[thick,<-] (-1/2,0) -- (-3/2,0);
	\node[below,scale=0.75] at (-1,0) {$h_p$};
	\draw[thick] (-3/2,0) -- (-3/2,0.8);
	\node[above,scale=0.75] at (-3/2,0.8) {$h_{i_2}$};
        \draw[thick] (-3/2,0) -- (-5/2,0);
        \node[left,scale=0.75] at (-5/2,0) {$h_{i_1}$};
	\end{tikzpicture}
	}} \right]^\dagger {\cal B}\left[  \vcenter{\hbox{\vspace{0.12in}
	\begin{tikzpicture}[scale=.75]
	\draw[thick,<-] (-1/2,0) -- (-3/2,0);
	\node[below,scale=0.75] at (-1,0) {$h_p$};
	\draw[thick] (-3/2,0) -- (-3/2,0.8);
	\node[above,scale=0.75] at (-3/2,0.8) {$h_{i_3}$};
        \draw[thick] (-3/2,0) -- (-5/2,0);
        \node[left,scale=0.75] at (-5/2,0) {$h_{i_4}$};
	\end{tikzpicture}
	}} \right] |p\rangle  \ ,
\end{align}
and similarly for $n$-point functions.

\subsection{The tensor network}
\renewcommand{\S}{\mathcal{S}}
Before we describe how to build the tensor network, let us briefly discuss what \textit{doesn't} work. Suppose we follow the same procedure that worked with spherical symmetry: Insert a complete set of energy eigenstates between each operator, and declare the resulting matrix product state to be a tensor network. This fails --- it's a valid CFT calculation, but the tensor ranks in this `eigenstate network' do not match the bulk geometry. For example, the tensor rank is not minimized at the link corresponding to a bulk minimal surface. The problem is that it does not make sense to treat the eigenstate network as pseudorandom. In a CFT with Virasoro symmetry, it is only the primary OPE coefficients that can plausibly be pseudorandom, not all of the matrix elements. Since the tensors in the eigenstate network are not pseudorandom, there is no reason to expect its entanglement structure to be simply related to the tensor geometry. The same comments apply to more general bulk theories whenever matter fields are turned on: the matrix elements of light, single-trace operators dual to weakly interacting fields in the bulk cannot be random.

To circumvent this, we need to use large $c$ and make one further assumption: that the conformal block expansion for $\langle \Psi | \Psi\rangle$ is dominated by a semiclassical saddlepoint with all of the internal weights above the black hole threshold, $h, \bh> \frac{c}{24}$. Denote these saddlepoint weights by $h_{k}^*$ for $k = 1,\dots,m-1$, labeled as follows:\footnote{ For  reflection-positive correlators which are dominated by the identity block, the saddlepoint weights are scalars, $h_k^* = \bh_k^*$. This follows from the fusion transformations described in \cite[section 8]{Chandra:2022bqq}: Starting from the identity block in the dual channel where operators fuse in conjugate pairs, the comb channel only has contributions with  $(h_i, \bh_i) = (h_{2m-i}, \bh_{2m-i})$;  reflection positive kinematics implies $(h_i, \bh_i) = (\bh_{2m-i}, h_{2m-i})$; therefore $h_i = \bh_i$.}
\begin{align}\label{saddleweights}
   \vcenter{\hbox{\vspace{0.12in}
	\begin{tikzpicture}[scale=.75]
	\draw[thick] (11/2,0) -- (13/2,0);
        \node[right,scale=0.75] at (13/2,0) {$i_{2m}$};
        \draw[thick] (11/2,0) -- (11/2,0.8);
        \node[above,scale=0.75] at (11/2,0.8) {$i_{2m-1}$};
        \draw[thick] (9/2,0) -- (11/2,0);
        \node[below, scale=0.75] at (5,0) {$h_1^*$};
        \draw[dashed] (7/2,0) -- (9/2,0);
        \draw[thick] (5/2,0) -- (7/2,0);
        \node[above, scale=0.75] at (5/2,0.8) {$i_{m+1}$};
        \draw[thick] (5/2,0) -- (5/2,0.8);
        \draw[thick] (3/2,0) -- (5/2,0);
        \node[below, scale=0.75] at (3.2,0) {$h_{m-2}^*$};
        \draw[thick] (3/2,0) -- (3/2,0.8);
        \node[above, scale=0.75] at (3/2,0.8) {$i_m$};
        \draw[thick] (1/2,0) -- (3/2,0); 
        \node[below, scale=0.75] at (2.2,0) {$h_{m-1}^*$};
	\draw[dashed] (-1/2,0) -- (1/2,0);
	\node[below,scale=0.75] at (1,0) {$h_{m-2}^*$};
	\draw[thick] (-1/2,0) -- (-3/2,0);
	\node[below,scale=0.75] at (-1,0) {$h_1^*$};
	\draw[thick] (-3/2,0) -- (-3/2,0.8);
	\node[above,scale=0.75] at (-3/2,0.8) {$i_2$};
        \draw[thick] (-3/2,0) -- (-5/2,0);
        \node[left,scale=0.75] at (-5/2,0) {$i_1$};
	\end{tikzpicture}
	}}
\end{align}
The fact that this correlation function is a norm $\langle \Psi | \Psi\rangle$ guarantees that the saddlepoint weights are symmetric across the diagram, as written. For fixed kinematics and external weights, let $\H_k$ be the set of primaries within a microcanonical window near the semiclassical saddlepoint $(h_k^*, \bh_k^*)$. The number of such states is given by the Cardy formula,
\begin{align}
|\H_k| \approx  e^{S_0(h_k^*, \bh_k^*)} \ , \quad
S_0(h,\bh) = 2\pi \sqrt{\frac{c}{6}(h - \frac{c}{24})} + 2 \pi \sqrt{ \frac{c}{6}(\bh - \frac{c}{24})}
\ .
\end{align}
Note that $\H_k$ is defined to include only primaries, not all states, but at large $c$ this doesn't affect the Cardy formula.

Now we define a semiclassical state by truncating the sums in \eqref{Bexp} to primaries near the saddlepoint,
\begin{align}\label{psistarexp}
|\Psi\rangle_* &:= \sum_{p_k \in \H_k}
 c_{i_1 i_2 p_1} \cdots c_{i_m p_{m-2} p_{m-1}} 
  \left| \B\left[ \vcenter{\hbox{\vspace{0.12in}
	\begin{tikzpicture}[scale=.75]
        \draw[thick,->] (3/2,0) -- (5/2,0);
        \node[below,scale=0.75] at (2,0) {$h_{p_{m-1}}$};
        \draw[thick] (3/2,0) -- (3/2,0.8);
        \node[above,scale=0.75] at (3/2,0.8) {$h_{i_m}$};
        \draw[thick] (1/2,0) -- (3/2,0); 
	\draw[dashed] (-1/2,0) -- (1/2,0);
	\node[below,scale=0.75] at (1,0) {$h_{p_{m-2}}$};
	\draw[thick] (-1/2,0) -- (-3/2,0);
	\node[below,scale=0.75] at (-1,0) {$h_{p_1}$};
	\draw[thick] (-1/2,0) -- (-1/2,1*0.8);
	\node[above,scale=0.75] at (-1/2,0.8) {$h_{i_3}$};
	\draw[thick] (-3/2,0) -- (-3/2,0.8);
	\node[above,scale=0.75] at (-3/2,0.8) {$h_{i_2}$};
        \draw[thick] (-3/2,0) -- (-5/2,0);
        \node[left,scale=0.75] at (-5/2,0) {$h_{i_1}$};
	\end{tikzpicture}
	}} \right] \right|^2|p_{m-1}\rangle  
 \end{align}
 The OPE block depends only on the weights, not the particular operators, so it can be evaluated at the saddlepoint and moved outside the sum. Therefore we obtain
 \begin{align}\label{psistarexp2}
|\Psi\rangle_* 
&= |\B|_*^2 \sum_{p_k \in \H_k}
 c_{i_1 i_2 p_1} \cdots c_{i_m p_{m-2} p_{m-1}}  |p_{m-1}\rangle
\end{align}
where
\begin{align}
|\B|_*^2 = \left| \B\left[\vcenter{\hbox{\vspace{0.12in}
	\begin{tikzpicture}[scale=.75]
        \draw[thick,->] (3/2,0) -- (5/2,0);
        \node[below,scale=0.75] at (2,0) {$h^*_{m-1}$};
        \draw[thick] (3/2,0) -- (3/2,0.8);
        \node[above,scale=0.75] at (3/2,0.8) {$h_{i_m}$};
        \draw[thick] (1/2,0) -- (3/2,0); 
	\draw[dashed] (-1/2,0) -- (1/2,0);
	\node[below,scale=0.75] at (1,0) {$h^*_{m-2}$};
	\draw[thick] (-1/2,0) -- (-3/2,0);
	\node[below,scale=0.75] at (-1,0) {$h_1^*$};
	\draw[thick] (-1/2,0) -- (-1/2,1*0.8);
	\node[above,scale=0.75] at (-1/2,0.8) {$h_{i_3}$};
	\draw[thick] (-3/2,0) -- (-3/2,0.8);
	\node[above,scale=0.75] at (-3/2,0.8) {$h_{i_2}$};
        \draw[thick] (-3/2,0) -- (-5/2,0);
        \node[left,scale=0.75] at (-5/2,0) {$h_{i_1}$};
	\end{tikzpicture}
	}}\right] \right|^2  \ .
\end{align}

\begin{figure}[t]
\begin{center}
\begin{tikzpicture}[scale=0.75]
\draw (-0.8, -.05) -- (0.2,-.05);
\draw (-0.8, .05) -- (0.2,.05);
\begin{scope}
\clip (0.2,-1) rectangle (0.7,1);
\draw[black,fill=btensorcolor] (0.2,0) circle (0.5);
\end{scope}
\draw (0.2,-0.5) -- (0.2,0.5);
\draw (2,0) -- (2.6,0);
\draw (1.7,0) -- (1.7,1);
\draw (0.7,0) -- (1.7,0);
\draw[black,fill=ctensorcolor] (1.7,0) circle (0.3);
\draw [dashed] (2.8,0) -- (4,0);
\draw (4.3,0) -- (4.3,1);
\draw (4.3,0) -- (5.3,0);
\draw[black,fill=ctensorcolor] (4.3,0) circle (0.3);
\draw (5.6,0) -- (5.6,1);
\draw (5.6,0) -- (6.6,0);
\draw[black,fill=ctensorcolor] (5.6,0) circle (0.3);
\node[above] at (4.3,1) {$i_3$};
\node[above] at (5.6,1) {$i_2$};
\node[right] at (6.6,0) {$i_1$};
\node[above] at (1.7,1) {$i_m$};
\node[left] at (-1.5,0) {$\ket{\Psi}_*=$};
\end{tikzpicture}
\end{center}
\caption{The semiclassical CFT state as a tensor network, as in \eqref{psistarexp2}. Internal lines are labeled by CFT primaries with $h,\bh > \frac{c}{24}$. The red tensors are OPE coefficients and the final tensor on the left is the (mod-squared) OPE block, which is a map from the space of primaries to the physical 
CFT Hilbert space. 
\label{fig:psinetwork}
}
\end{figure}

The expression \eqref{psistarexp2} is manifestly in the form of a 1d tensor network with the architecture of a matrix product state. Denote the tensors as
\begin{align}\label{tensordefinitions}
c_{pqr} \quad &= \quad
\vcenter{\hbox{
\begin{tikzpicture}[scale=0.75]
\draw (-1,0) -- (1, 0);
\draw (0, 0) -- (0,1);
\node[left] at (-1,0) {\small $p$};
\node[above] at (0,1) {\small $q$};
\node[right] at (1,0) {\small $r$};
\draw[black,fill=ctensorcolor] (0,0) circle (0.3);
\end{tikzpicture}
}}
\\
\langle m| |\B|_*^2 |p\rangle \quad &= \quad
\vcenter{\hbox{
\begin{tikzpicture}[scale=0.75]
\draw (0,0) -- (1, 0);
\draw (1.5, -.05) -- (2.5,-.05);
\draw (1.5, .05) -- (2.5,.05);
\draw[thick] (1.5,-0.5) -- (1.5,0.5);
\node[left] at (0,0) {\small $p$};
\node[right] at (2.5,0) {\small $m$};
\begin{scope}
\clip (1,-1) rectangle (1.5,1);
\draw[black,fill=btensorcolor] (1.5,0) circle (0.5);
\end{scope}
\end{tikzpicture}
}}
\end{align}
Single lines are tensor legs that act in the space of CFT primaries, $\H_{\rm prim}$. Double lines act in the physical Hilbert space, $\H_{\rm CFT}$. The semiclassical OPE block is a linear map 
\begin{align}
|\B|_*^2: \H_{\rm prim} \to \H_{\rm CFT} \ ,
\end{align}
so the corresponding tensor has one leg with a single line, and one leg with a double line. In this notation, the semiclassical state \eqref{psistarexp2} is the tensor network in fig.~\ref{fig:psinetwork}.

 The sum in $|\Psi\rangle_*$ is by definition truncated to primaries near the saddlepoint, so the internal tensor legs in fig.~\ref{fig:psinetwork} have finite dimension. The truncation was essential in order to write the state as a tensor network --- otherwise, we cannot extract the OPE block outside the sum in \eqref{psistarexp2}. Thus the exact CFT state is not a tensor network with the architecture of figure \ref{fig:psinetwork}. This clarifies the sense in which we should expect holographic geometries to be similar to random tensor networks.  (It also suggests that more general, non-holographic CFT states can nonetheless be similar to holographic tensor networks in a regime where the conformal block expansion is dominated by a saddlepoint.)

Now the goal is to understand how the tensor network in fig.~\ref{fig:psinetwork} discretizes the radial direction in the bulk.

\subsection{The random tensor approximation}
There is evidence that semiclassical 3d gravity is holographically dual to an ensemble of large-$c$ CFTs \cite{Cotler:2020ugk,Belin:2020hea,Chandra:2022bqq}. In \cite{Belin:2020hea,Chandra:2022bqq}, the ensemble is defined by treating the primary OPE coefficients as random variables. At leading order,\footnote{This is conjectured to be `leading' in the sense that corrections on the right-hand side of \eqref{censemble} come with factors of $e^{-S}$. This does not always mean that the terms in \eqref{censemble} give the leading contribution to observables, though in many cases it does. We have also assumed for simplicity that all three operators have $h > \frac{c}{32}$ to avoid complications from multi-twist operators, which have OPE coefficients determined by Virasoro \cite{Collier:2018exn}, and that they are heavy enough to support a 3-point wormhole.
See \cite{Belin:2020hea,Chandra:2022bqq} for more discussion.}
the OPE coefficients are Gaussian random variables with   \cite{Chandra:2022bqq}
\begin{align}\label{censemble}
\overline{c_{ijk}c^*_{lmn}} = C_0(h_i, h_j, h_k)C_0(\bh_i, \bh_j, \bh_k) \left( \delta_{il} \delta_{jm} \delta_{kn} \pm \mbox{permutations}\right) \ .
\end{align}
The coefficient $C_0$ is the crossing kernel for the Virasoro identity block; it is a smooth function of the weights that is complicated, but known explicitly \cite{Collier:2019weq}. This choice ensures that the CFT ensemble reproduces the identity block approximation in 3d gravity \cite{Hartman:2013mia,Faulkner:2013yia}. 
Thus \eqref{censemble}, by design, reproduces the correlation functions of conical defects and partition functions of handlebodies in AdS$_3$. Much more nontrivial is that \eqref{censemble} also matches the contribution of a wide variety of multi-boundary wormholes \cite{Chandra:2022bqq}.

The ansatz \eqref{censemble} is a version of the eigenstate thermalization hypothesis \cite{PhysRevA.43.2046,Srednicki:1994mfb}, tailored to holographic 2d CFTs. Combined with the tensor network representation of black hole pure states in fig.~\ref{fig:psinetwork}, the large-$c$ ensemble is naturally interpreted as a random tensor network model. The tensors are the primary OPE coefficients $c_{ijk}$, truncated to the finite set of states near the semiclassical saddlepoint. 

\subsection{Sphere 4-point functions}

As an example, consider the state 
\begin{align}
|\Psi_2\rangle = \O^2(-\tau_2, \phi_2) \O^1(-\tau_1, \phi_1) |0\rangle \ ,
\end{align}
where $\O^{1}$ and $\O^2$ are scalar primaries with $h \in \left( \frac{c}{32}, \frac{c}{24}\right)$. With weights in this range, all of the states in the OPE $\O^1 \O^2$ are black holes \cite{Collier:2018exn}. 
The corresponding tensor network is 
\begin{align}\label{network2}
|\Psi_2\rangle_* = \vcenter{\hbox{
\begin{tikzpicture}[scale=0.75]
\draw (1.5, -.05) -- (2.5,-.05);
\draw (1.5, .05) -- (2.5,.05);
\begin{scope}
\clip (2.5,-1) rectangle (3,1);
\draw[black,fill=btensorcolor] (2.5,0) circle (0.5);
\end{scope}
\draw (2.5,-0.5) -- (2.5,0.5);
\draw (3,0) -- (4,0);
\draw (4.3,0) -- (4.3,1);
\draw (4.3,0) -- (5.3,0);
\draw[black,fill=ctensorcolor] (4.3,0) circle (0.3);
\node[above] at (4.3,1) {$2$};
\node[right] at (5.3,0) {$1$};
\end{tikzpicture}
}}
\end{align}
The norm is the 4-point function,
\begin{align}
\langle \Psi_2 | \Psi_2 \rangle \approx {}_*\langle \Psi | \Psi \rangle_* =\vcenter{\hbox{
\begin{tikzpicture}[scale=0.75]
\draw (-1.3,0) -- (-0.3,0);
\draw (-0.3, 0) -- (-0.3,1);
\node[left] at (-1.3,0) { $1$};
\node[above] at (-0.3,1) {$2$};
\node[right] at (1,0) {\small $r$};
\draw[black,fill=ctensorcolor] (-0.3,0) circle (0.3);
\draw (0,0) -- (1, 0);
\draw (1.5, -.05) -- (2.5,-.05);
\draw (1.5, .05) -- (2.5,.05);
\draw[thick] (1.5,-0.5) -- (1.5,0.5);
\begin{scope}
\clip (1,-1) rectangle (1.5,1);
\draw[black,fill=btensorcolor] (1.5,0) circle (0.5);
\end{scope}
\begin{scope}
\clip (2.5,-1) rectangle (3,1);
\draw[black,fill=btensorcolor] (2.5,0) circle (0.5);
\end{scope}
\draw (2.5,-0.5) -- (2.5,0.5);
\draw (3,0) -- (4,0);
\draw (4.3,0) -- (4.3,1);
\draw (4.3,0) -- (5.3,0);
\draw[black,fill=ctensorcolor] (4.3,0) circle (0.3);
\node[above] at (4.3,1) {$2$};
\node[right] at (5.3,0) {$1$};
\end{tikzpicture}
}}
\end{align}
The identity \eqref{pFFp} in tensor network notation is 
\begin{align}
\vcenter{\hbox{
\begin{tikzpicture}[scale=0.75]
\draw (0,0) -- (1, 0);
\node[left] at (0,0) {$p$};
\draw (1.5, -.05) -- (2.5,-.05);
\draw (1.5, .05) -- (2.5,.05);
\draw[thick] (1.5,-0.5) -- (1.5,0.5);
\begin{scope}
\clip (1,-1) rectangle (1.5,1);
\draw[black,fill=btensorcolor] (1.5,0) circle (0.5);
\end{scope}
\begin{scope}
\clip (2.5,-1) rectangle (3,1);
\draw[black,fill=btensorcolor] (2.5,0) circle (0.5);
\end{scope}
\draw (2.5,-0.5) -- (2.5,0.5);
\draw (3,0) -- (4,0);
\node[right] at (4,0) {$q$};
\end{tikzpicture}
}}= |{\cal F}|^2 \delta_{pq}
\end{align}
where ${\cal F}$ is the 4-point conformal block.\footnote{Note that the primaries in $|\Psi\rangle_*$ are truncated to near the saddlepoint, but all descendants are retained, so this formula is exact.} Therefore we can also write the norm as a tensor network of OPE coefficients, weighted by conformal blocks:
\begin{align}
\langle \Psi_2|\Psi_2\rangle \approx  \sum \vcenter{\hbox{
\begin{tikzpicture}[scale=0.75]
\draw (-1.3,0) -- (-0.3,0);
\draw (-0.3, 0) -- (-0.3,1);
\node[left] at (-1.3,0) { $1$};
\node[above] at (-0.3,1) {$2$};
\draw[black,fill=ctensorcolor] (-0.3,0) circle (0.3);
\draw (0,0) -- (1, 0);
\draw (1.3,0) -- (1.3,1);
\draw (1.3,0) -- (2.3,0);
\draw[black,fill=ctensorcolor] (1.3,0) circle (0.3);
\node[above] at (1.3,1) {$2$};
\node[right] at (2.3,0) {$1$};
\end{tikzpicture}
}} |{\cal F}|^2 \ .
\end{align}
This is almost the usual conformal block expansion, in different notation --- the diagram represents the product of OPE coefficients, see \eqref{tensordefinitions}.  It is not quite the usual conformal block expansion, however, because the tensors are by definition truncated near the semiclassical saddle, and have finite dimension.
 
The log-dimension of the internal leg in \eqref{network2} is $S_0(h_1^*, \bh_1^*)$, where $(h_1^*, \bh_1^*)$ are the conformal weights of the primary that dominates the conformal block sum. We call this the \textit{primary entropy} --- the Cardy entropy of the lowest weight state in the representation. Generally, asymmetric excited states can have leading-order contributions to the energy from Virasoro descendants, so it is important to remove them before applying the Cardy formula, and $(h_1^*, \bh_1^*)$ differ at leading order from the total conformal weights at the saddlepoint.

The geometry dual to $|\Psi_2\rangle_*$ is a black hole created by two conical defects. The metric is given in \eqref{BHMetric} below. The $t=0$ time-symmetric spatial slice looks schematically like this:
\begin{align}\label{spiky2}
    \vcenter{\hbox{
\begin{overpic}[width=1.5in,grid=false]{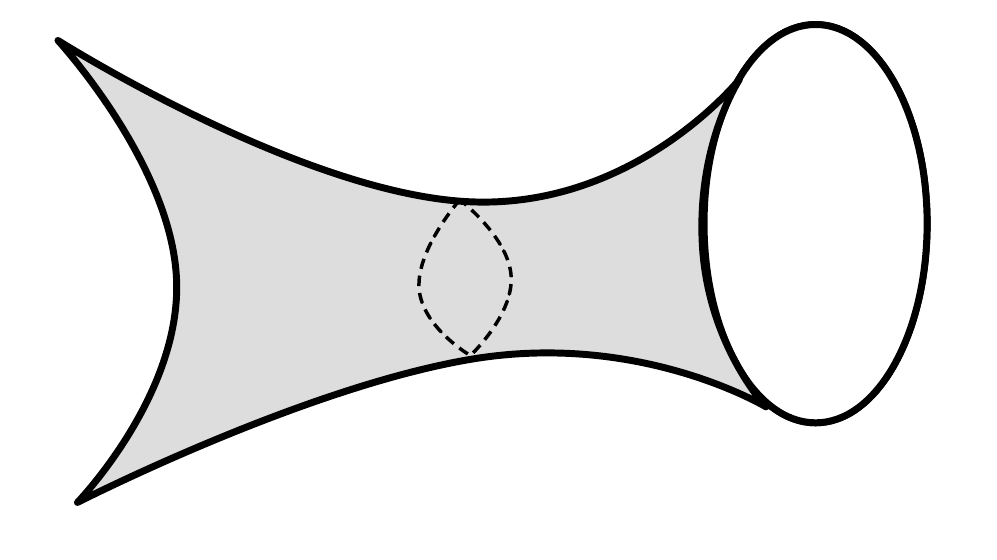}
\put (47,10) {{$\gamma$}}
\end{overpic}
}}
\end{align}
The locally-minimal surface $\gamma$ is a time-symmetric apparent horizon. The area of this apparent horizon is interpreted as a coarse-grained entropy \cite{Engelhardt:2017aux,Chandra:2022fwi}. In 3d gravity, the area is related to the primary entropy in the OPE:
\begin{align}
S_0(h_1^*, \bh_1^*) = \frac{\mbox{Area}(\gamma)}{4} \ . 
\end{align}
(In three bulk dimensions, `area' means length.) 
This will be derived from a gravity calculation in section \ref{s:isometric}. Note that the state $|\Psi_2\rangle$ is pure, so its von Neumann entropy is zero, in agreement with the Ryu-Takayanagi formula applies to the trivial (empty) surface.

Finally we can compare the tensor network  in \eqref{network2} to the bulk spatial slice in \eqref{spiky2}. The comparison is a bit trivial in this case, because the tensor network has only one internal line. But the two pictures agree: The red tensor corresponds to the black hole interior, the OPE block corresponds to the near-boundary region, and the log-dimension of the internal line is $\frac{1}{4}\mbox{Area}(\gamma)$.

\subsection{Probes of the apparent horizon}\label{ss:probe6}
The 6-point function is more interesting, because here we can study the isometric property and the breakdown of the Virasoro identity approximation. Consider 
\begin{align}
|\Psi_3\rangle = \O^3(-\tau_3, \phi_3) \O^2(-\tau_2,\phi_2) \O^1(-\tau_1,\phi_1)|0\rangle \ .
\end{align}
The tensor network is
\begin{align}\label{threeopTN}
|\Psi_3\rangle_*  = \vcenter{\hbox{
\begin{tikzpicture}[scale=0.75]
\draw (1.5, -.05) -- (2.5,-.05);
\draw (1.5, .05) -- (2.5,.05);
\begin{scope}
\clip (2.5,-1) rectangle (3,1);
\draw[black,fill=btensorcolor] (2.5,0) circle (0.5);
\end{scope}
\draw (2.5,-0.5) -- (2.5,0.5);
\draw (3,0) -- (4,0);
\draw (4.3,0) -- (4.3,1);
\draw (4.3,0) -- (5.3,0);
\draw[black,fill=ctensorcolor] (4.3,0) circle (0.3);
\draw (5.6,0) -- (5.6,1);
\draw (5.6,0) -- (6.6,0);
\draw[black,fill=ctensorcolor] (5.6,0) circle (0.3);
\node[above] at (4.3,1) {$3$};
\node[above] at (5.6,1) {$2$};
\node[right] at (6.6,0) {$1$};
\node[below] at (5,0) {$h_1^*$};
\node[below] at (3.5,0) {$h_2^*$};
\end{tikzpicture}
}}
\end{align}
There are now two internal lines, with saddlepoint primary weights $h_1^*$ and $h_2^*$.  The log-dimensions of these tensor legs are equal to the primary entropies,
\begin{align}\label{sixpointweights}
S_1 = S_0(h_1^*, \bh_1^*) \quad \mbox{and} \quad S_2 = S_0(h_2^*, \bh_2^*) \ .
\end{align}
In the random tensor approximation, the tensor corresponding to the OPE coefficient $c_{3pq}$ (with $p$ and $q$ the internal states corresponding to $h_1^*$ and $h_2^*$ respectively) is a rectangular random matrix, with log-dimensions given by \eqref{sixpointweights}. Effectively, it is a map from the primary Hilbert space around $h_1^*$ to the primary Hilbert space around $h_2^*$, which we denote\footnote{The exact OPE coefficient is of course infinite-dimensional. Here by $c_{3pq}$ we mean the associated tensor in the tensor network, which is by definition truncated to states near the semiclassical saddle. 
}
\begin{align} 
c_{3pq} : \H_1 \to \H_2 \ .
\end{align}
Therefore, this map is approximately isometric or co-isometric, depending on the relative size of the input and output spaces:
\begin{align}\label{shier}
S_1 < S_2 \quad &\Rightarrow \quad c_{3pq} \mbox{\ isometric} \\
S_1 > S_2 \quad &\Rightarrow \quad c_{3pq} \mbox{\ co-isometric} \notag
\end{align}
Let us suppose $h_3 \ll c$, so that $\O^3$ is a probe operator; the dual particle travels on a spacelike geodesic in the background black hole created by $O^2O^1$. In section \ref{s:isometric}, we show that the two cases in \eqref{shier} correspond to whether the probe particle is outside or inside the apparent horizon. If the probe particle sits exactly on the horizon, then the primary energies in the saddlepoint OPE are equal, $h_1^*=h_2^*$. This does not hold for the total energy, only the primary energy, so it was essential that we built the network using the auxiliary Hilbert space  $\H_{\rm prim}$.

If $h_3/c$ is finite, then $\O^3$ backreacts, and there is no longer any simple relation between the isometric property of the random map $c_{3pq}$ and whether or not $\O^3$ is behind the horizon. The marginal case $S_1 = S_2$ defines a natural notion of non-perturbative horizon that is explored in section \ref{nonperp}.  
The relationship between the isometric property and the apparent horizon carries over to other observables, including higher-point functions, the BTZ black hole, and BTZ plus heavy particles. All of these cases are analyzed in sections \ref{s:isometric}-\ref{s:btz}. 

\subsection{Reconstruction of simple operators}

We normalise the random matrix $c_{3pq}$ (again, truncated to states near the saddle) by 
\begin{equation}
    V_{pq}=\frac{1}{e^{S_2/2}}\frac{c_{3pq}}{\sqrt{C_0(h_3,h_1^*,h_2^*)C_0(h_3,\overline{h}_1^*,\overline{h}_2^*)}}
\end{equation}
Note that $Ve^{S_2/2}$ is a complex Gaussian random matrix with vanishing mean and unit variance. In the semiclassical limit, $V$ always acts isometrically on average irrespective of whether the probe goes inside or outside the horizon i.e,
\begin{equation}
    \overline{ V^{\dagger}V } =\id_{\mathcal{H}_1}
\end{equation}
with the average taken over the space of normalised complex Gaussian random matrices with the usual measure. For the case where the probe is outside the horizon, the stronger statement that $V^{\dagger}V = \id_{\mathcal{H}_1}$ holds (at leading order).  This cannot be true for the case where the probe goes inside the horizon simply because $\text{rank}(V^{\dagger}V)\leq \text{min}(e^{S_1},e^{S_2})$. However, a typical $V$ still preserves the overlaps of a large number of states in $\mathcal{H}_1$. For instance, one can show that for a randomly chosen $V$ from the ensemble,
\begin{equation} \label{Devbound}
    \text{Pr}\left(\bigg|||V\ket{\psi}||-1\bigg|\geq e^{-\alpha S_2}\right) \leq 2e^{-\frac{1}{2}e^{(1-2\alpha) S_2}}
\end{equation}
for any $\ket{\psi}\in \mathcal{H}_1$ normalised so that $\langle \psi \ket{\psi}=1$. Here, $\alpha$ is a parameter which takes values in the range $(0,\frac{1}{2})$. This result says that the probability for the norm of any normalised state in $\mathcal{H}_1$ to deviate from unity by more than an exponentially small quantity in $S_2$ is doubly exponentially small in $S_2$. In other words, even when the code is non-isometric, it is very likely to preserve the norm of any particular state. This explains how the bulk effective field theory can still provide a good description of the black hole interior for many purposes \cite{Akers:2022qdl,Kar:2022qkf}. The derivation of this result is similar to that presented in \cite[section 3]{Kar:2022qkf} so we refer the reader to \cite{Kar:2022qkf} for details.

\subsection{Breakdown of the identity block approximation}

This analysis can be re-phrased in terms of a subtle breakdown of the Virasoro identity block approximation.  Consider the 6-point function,
\begin{align}
G_6 = \langle \O^{1\dagger} \O^{2\dagger} \O^{3\dagger} \O^3 \O^2 \O^1 \rangle
\end{align}
which is the norm-squared of the state  $ |\Psi_3\rangle$ considered in the previous subsection. In a holographic CFT, this 6-point function is computed in the bulk by a geometry with three conical defects,
\begin{align}
G_6 &\approx  \vcenter{\hbox{
\begin{overpic}[width=1.5in,grid=false]{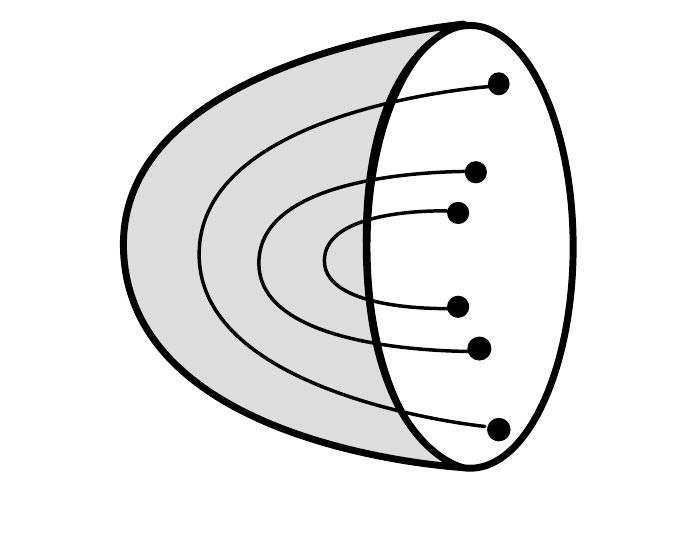}
\end{overpic}
}}
\end{align}
As demonstrated in \cite{Hartman:2013mia,Faulkner:2013yia}, the on-shell action of the semiclassical saddlepoint is reproduced by a large-$c$ Virasoro identity block in the channel where the operators fuse in pairs:
\begin{align}
G_6 &\approx \left|\vcenter{\hbox{
	\begin{tikzpicture}[scale=0.75]
	\draw[thick] (0,0) -- (0,1);
	\draw[thick](0,1) -- (-0.866*0.8,1+0.8*0.5);
	\draw[thick] (0,1) -- (0.866*0.8,1+0.8*0.5);
	\draw[thick] (0,0) -- (0.866,-1/2);
	\draw[thick] (0,0) -- (-0.866,-1/2);
	\draw[thick] (0.866,-1/2) -- (0.866+0.8*0.866,-1/2+0.8*1/2);
	\draw[thick] (0.866,-1/2) -- (0.866,-1/2-0.8);
	\draw[thick] (-0.866,-1/2) -- (-0.866-0.8*0.866,-1/2+0.8*1/2);
	\draw[thick] (-0.866,-1/2) -- (-0.866,-1/2-0.8);
	\node[left,scale=0.75] at (-0.866*0.8,1+0.8*0.5) {$2$};
	\node[right,scale=0.75] at (0.866*0.8,1+0.8*0.5) {$2$};
	\node[right,scale=0.75] at (0.866+0.8*0.866,-1/2+0.8*1/2) {$3$};
	\node[below,scale=0.75] at (0.866,-1/2-0.8) {$3$};
	\node[left, scale=0.75] at (-0.866-0.8*0.866,-1/2+0.8*1/2) {$1$};
	\node[below, scale=0.75] at (-0.866,-1/2-0.8) {$1$};
	\node[left, scale=0.75] at (0,1/2) {$\id$};
	\node[above, scale=0.75] at (1/2*0.866,-1/2*1/2) {$\id$};
	\node[above, scale=0.75] at (-1/2*0.866,-1/2*1/2) {$\id$};
	\end{tikzpicture}
	}}  \right |^2
\end{align}
By a sequence of fusion moves (see \cite{Anous:2020vtw} and \cite[section 8.2]{Chandra:2022bqq}) this is equivalent to
\begin{align}
G_6 &\approx \left |\int_{c-1\over 24}^\infty dh_p\, 
	\rho_0(h_p)C_0(h_1,h_2,h_p)
	\vcenter{\hbox{
	\begin{tikzpicture}[scale=.75]
	\draw[thick] (0,0) -- (1,0);
	\node[above,scale=0.75] at (1/2,0) {$\id$};
	\draw[thick] (0,0) -- (0,1);
	\node[left,scale=0.75] at (0,1/2) {$p$};
	\draw[thick] (0,0) -- (0,-1);
	\node[left,scale=0.75] at (0,-1/2) {$p$};
	\draw[thick] (0,1) -- (0.8*0.866,1+0.8*1/2);
	\draw[thick] (0,1) -- (-0.8*0.866,1+0.8*1/2);
	\node[above,scale=0.75] at (-0.8*0.866,1+0.8*1/2) {$1$};
	\node[above,scale=0.75] at (0.8*0.866,1+0.8*1/2) {$2$};
	\draw[thick] (0,-1) -- (0.8*0.866,-1-0.8*1/2);
	\draw[thick] (0,-1) -- (-0.8*0.866,-1-0.8*1/2);
	\node[below,scale=0.75] at (-0.8*0.866,-1-0.8*1/2) {$1$};
	\node[below,scale=0.75] at (0.8*0.866,-1-0.8*1/2) {$2$};
	\draw[thick] (1,0) -- (1+0.8*1/2,0.8*0.866);
	\draw[thick] (1,0) -- (1+0.8*1/2,-0.8*0.866);
	\node[right,scale=0.75] at (1+0.8*1/2,0.8*0.866) {$3$};
	\node[right,scale=0.75] at (1+0.8*1/2,-0.8*0.866) {$3$};
	\end{tikzpicture}
	}}\right |^2
\end{align}
which corresponds to the OPE coefficients given in \eqref{censemble}. 
In the identity approximation, the only operators retained in the $\O^{3\dagger}\O^3$ OPE are the identity operator and its descendants. Therefore $\O^{3\dagger}\O^3$ acts diagonally on the primary labels in this approximation:
\begin{align}\label{o3id}
\langle p, N, \bar{N}| \O^{3\dagger} \O^3 | p', M, \bar{M}\rangle \propto \delta_{pp'} \ , 
\end{align}
where $p,p'$ label Virasoro representations, and $N, \bar{N}, M, \bar{M}$ label descendants. 

We must distinguish between a strong version of the identity approximation, in which $\O^{3\dagger}\O^3$ acts diagonally on representations in the sense of an operator, and a weak version, where \eqref{o3id} only holds element-by-element in the primary basis. 
According to the discussion above, the strong version is impossible if $\O^3$ is dual to a probe particle behind the apparent horizon. Consider the tensor operator
\begin{align}
K =\vcenter{\hbox{
\begin{tikzpicture}[scale=0.75]
\draw (-1.3,0) -- (-0.3,0);
\draw (-0.3, 0) -- (-0.3,1);
\node[above] at (-0.3,1) {$3$};
\draw[black,fill=ctensorcolor] (-0.3,0) circle (0.3);
\draw (0,0) -- (1, 0);
\draw (1.5, -.05) -- (2.5,-.05);
\draw (1.5, .05) -- (2.5,.05);
\draw[thick] (1.5,-0.5) -- (1.5,0.5);
\begin{scope}
\clip (1,-1) rectangle (1.5,1);
\draw[black,fill=btensorcolor] (1.5,0) circle (0.5);
\end{scope}
\begin{scope}
\clip (2.5,-1) rectangle (3,1);
\draw[black,fill=btensorcolor] (2.5,0) circle (0.5);
\end{scope}
\draw (2.5,-0.5) -- (2.5,0.5);
\draw (3,0) -- (4,0);
\draw (4.3,0) -- (4.3,1);
\draw (4.3,0) -- (5.3,0);
\draw[black,fill=ctensorcolor] (4.3,0) circle (0.3);
\node[above] at (4.3,1) {$3$};
\end{tikzpicture}
}}
\end{align}
which maps $\H_1 \to \H_1$. The relation \eqref{o3id}, in the strong sense, would imply $K \propto \id_{\H_1}$ as an operator. However, the rank of this operator is bounded above by the dimension of the internal line,
\begin{align}
\mbox{rank}\, K \leq \dim \H_2 \ .
\end{align}
Therefore $K$ cannot be proportional to the identity when $S_2 < S_1$ and the code is co-isometric --- its rank is less than its dimension. It is still possible for $\langle p | K | q \rangle \approx \delta_{pq}$ in the weak sense that it holds for individual matrix elements in the primary basis, but it cannot hold in the strong sense that $K \propto \id_{\H_1}$ as an operator. 

The conclusion is that when $\O^3$ is dual to a probe inside the horizon, the identity approximation must fail at \textit{leading} order in sufficiently complicated interior states.
That is, $\O_3^\dagger \times \O_3 \approx \mathbb{1}_{\rm Vir}$ is a good approximation in states created by a small number of primary operators, but there must exist superpositions of the form
\begin{align}
|a\rangle = \sum_{i,j,k,\dots} a_{ijk\dots} \O_i \O_j \O_k\dots|0\rangle
\end{align}
in which 
\begin{align}
\langle a| \O_3^\dagger \O_3| a \rangle
\end{align}
is not well approximated by the identity block.

 This distinction explains how the Virasoro identity block approximation in the CFT can reproduce the bulk EFT calculation of $G_6$ even when the operator is behind the horizon --- this is a low-energy observable that only depends on the weak identity approximation, not the strong one. The strong identity approximation breaks down at the horizon, precisely when the code transitions from isometric to co-isometric. This is the same behavior observed in the qubit models in \cite{Akers:2022qdl}.

\subsection{Tensor networks with higher topology}

\subsubsection{Thermal 2-point functions}\label{sss:btznetwork}

The above discussion readily generalises to states in several copies of the CFT Hilbert space. Consider for example the following state in two copies of the CFT Hilbert space obtained by exciting the thermofield double by a local operator,
\begin{equation} \label{twocopystate}
  \ket{\Psi}= \mathcal{O}\ket{\text{TFD}}=\sum_{p,q} c_{\mathcal{O}pq} \left | {\cal B}\left[  \vcenter{\hbox{\vspace{0.12in}
	\begin{tikzpicture}[scale=.75]
	\draw[thick,<-] (-1/2,0) -- (-3/2,0);
	\node[below,scale=0.75] at (-1,0) {$h_p$};
	\draw[thick] (-3/2,0) -- (-3/2,0.8);
	\node[above,scale=0.75] at (-3/2,0.8) {$h_{\mathcal{O}}$};
        \draw[thick,->] (-3/2,0) -- (-5/2,0);
        \node[below,scale=0.75] at (-2,0) {$h_q$};
	\end{tikzpicture}
	}} \right] \right|^2 \ket{p}\ket{q}
\end{equation}
where $\mathcal{O}$ is a scalar primary operator below the black hole threshold, and this OPE block is defined by reorganizing the sum over all states on the left-hand side into Virasoro representations. This is known as a partially entangled thermal state (PETS), and it has been studied in the SYK model and 2d gravity in \cite{Goel:2018ubv,Lin:2022rzw}. The 3d bulk dual of the above PETS state in 2d CFT is constructed in section \ref{PETSgeo}. The norm of this state computes the thermal two-point function $\langle\mathcal{O}\mathcal{O}\rangle_{\beta}$ and can be expanded using torus two-point conformal blocks,\footnote{There is implicit, nontrivial dependence on the operator location in \eqref{twocopystate}, that enters through the definition of the OPE block. Both operators are inserted at the same point on the spatial circle, so that this 2-point function may be viewed as a norm.}
\begin{equation} \label{twocopystatenorm}
   \langle \Psi \ket{\Psi}= \sum_{p,q} \left| c_{Opq} \right|^2 \left|\vcenter{\hbox{
	\begin{tikzpicture}[scale=0.75]
	\draw[thick] (0,0) circle (1);
	\draw[thick] (-1,0) -- (-2,0);
	\node[left] at (-2,0) {$\mathcal{O}$};
	\node[above] at (0,1) {$p$};
	\node[below] at (0,-1) {$q$};
	\draw[thick] (1,0) -- (2,0);
	\node[right] at (2,0) {$\mathcal{O}$};
	\end{tikzpicture}
	}} \right|^2
\end{equation}
Truncating the sum in (\ref{twocopystate}) around the semiclassical saddlepoint in the norm (\ref{twocopystatenorm}), we get the corresponding tensor network, 
\begin{equation} \label{twocopyTN}
    \ket{\Psi}_*=\vcenter{\hbox{
\begin{tikzpicture}[scale=0.75]
\draw (3,0) -- (4,0);
\node[left] at (3,0) {$\mathcal{O}$};
\draw (4.3,0) -- (4.3+0.5,0.886);
\draw (4.3,0) -- (4.3+0.5,-0.886);
\draw (4.3+0.5,0.886) -- (5.8,0.886);
\draw (4.3+0.5,-0.886) -- (5.8,-0.886);
\draw[black,fill=ctensorcolor] (4.3,0) circle (0.3);
\node[above] at (5.3,-0.886) {$h_R^*$};
\node[below] at (5.3,0.886) {$h_L^*$};
\draw[black,fill=btensorcolor] (5.8,0.4) rectangle (7,1.4);
\draw[black,fill=btensorcolor] (5.8,-0.4) rectangle (7,-1.4);
\draw (7,0.936) -- (8,0.936);
\draw (7,0.836) -- (8,0.836);
\draw (7,-0.836) -- (8,-0.836);
\draw (7,-0.936) -- (8,-0.936);
\node[right] at (5.65,0.9) {$|\mathcal{B}_L|^2_*$};
\node[right] at (5.65,-0.9) {$|\mathcal{B}_R|^2_*$};
\end{tikzpicture}
}}
\end{equation}
In the tensor network, the OPE block evaluated at the saddlepoint weights is interpreted as a map $|\mathcal{B}|^2_*: \mathcal{H}_{\text{prim}}\otimes \mathcal{H}_{\text{prim}}\to \mathcal{H}_{\text{CFT}}\otimes  \mathcal{H}_{\text{CFT}}$. The OPE block factorises between the two boundaries in the semiclassical limit so that $|\mathcal{B}|^2_*=|\mathcal{B}_L|^2_*|\mathcal{B}_R|^2_*$. This is because in the large-$c$ limit, the two-point function of the stress tensor evaluated on the two spatial boundaries factorises into a product of the semiclassical Liouville stress tensors (which solve the Liouville monodromy problem on the punctured cylinder with ZZ boundary conditions provided at either end) evaluated at the corresponding points. Since the connected contribution is subleading, this argument shows that the Virasoro excitations on the two boundaries are not entangled in the semiclassical limit, hence the OPE block factorises. This tensor network discretizes the $t=0$ spatial slice of BTZ backreacted with a conical defect. The defect may be inside or outside the horizon; if it is inside, then the $t=0$ spatial slice is
\begin{align}\label{btzdf}
    \vcenter{\hbox{
\begin{overpic}[width=1.8in,grid=false]{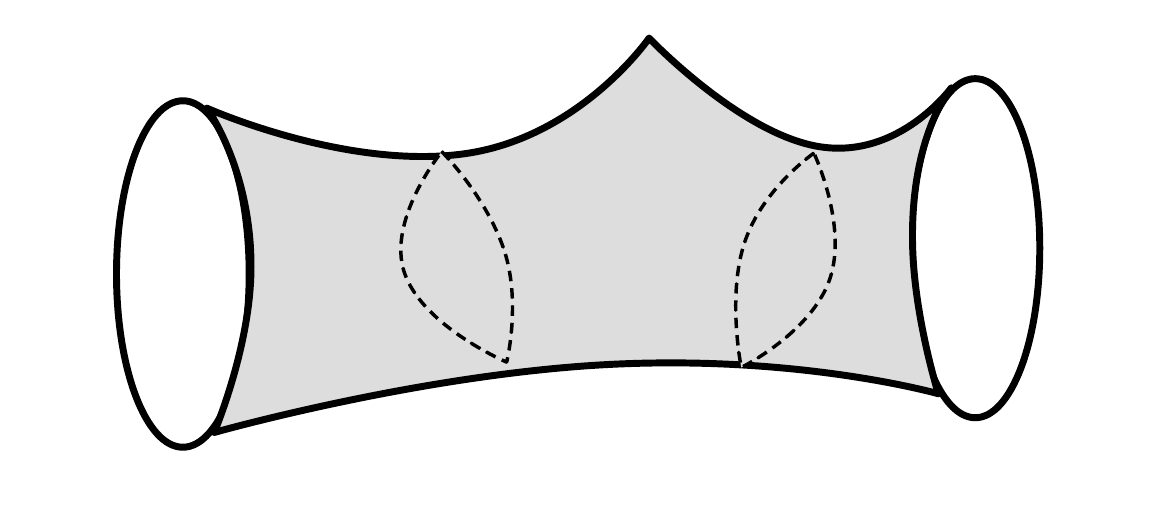}
\put (63,6) {$\gamma_R$}
\put (42,6)  {$\gamma_L$}
\end{overpic}
}}
\end{align}
In section \ref{s:btz}, we will show that the two time-symmetric apparent horizons indicated by $\gamma_L,\gamma_R$ in the figure above have areas matching with the primary entropy at the saddlepoint weights,
\begin{equation}
    S_0(h_R^*,\bh_R^*)=\frac{\text{Area}(\gamma_R)}{4}, \quad \quad \quad  S_0(h_L^*,\bh_L^*)=\frac{\text{Area}(\gamma_L)}{4}
\end{equation}
This matches with the bond dimensions of the corresponding internal legs in the network (\ref{twocopyTN}), so that we can view \eqref{twocopyTN} as a discretization of the radial direction in \eqref{btzdf}.

Depending on the location and weight of the operator ${\cal O}$, the defect may also be outside the BTZ horizon. From the CFT point of view one can distinguish these two possibilities by adding an additional probe particle and checking for an isometric/co-isometric transition. This is discussed in section \ref{s:btz} below.

\subsubsection{Genus-two partition functions}

Now, we give an example of a tensor network which discretizes the spatial slice of a smooth black hole geometry. Consider the state in three copies of the CFT Hilbert space,
\begin{equation} \label{threecopystate}
  \ket{\Psi}= \sum_{p,q,r} c_{pqr} \left | {\cal B}\left[  \vcenter{\hbox{\vspace{0.12in}
	\begin{tikzpicture}[scale=.75]
	\draw[thick,<-] (-1/2,0) -- (-3/2,0);
	\node[below,scale=0.75] at (-1,0) {$h_p$};
	\draw[thick,->] (-3/2,0) -- (-3/2,1);
	\node[right,scale=0.75] at (-3/2,0.5) {$h_{r}$};
        \draw[thick,->] (-3/2,0) -- (-5/2,0);
        \node[below,scale=0.75] at (-2,0) {$h_q$};
	\end{tikzpicture}
	}} \right] \right|^2 \ket{p}\ket{q}\ket{r}
\end{equation}
The norm of this state is the genus-two partition function,
\begin{equation} \label{threecopystatenorm}
   \langle \Psi \ket{\Psi}= Z_{g=2} = \sum_{p,q,r} \left| c_{pqr} \right|^2 \left|\vcenter{\hbox{
	\begin{tikzpicture}[scale=0.75]
	\draw[thick] (0,0) circle (1);
	\draw[thick] (-1,0) -- (1,0);
	\node[above] at (0,0) {$r$};
	\node[above] at (0,1) {$q$};
	\node[above] at (0,-1) {$p$};
	\end{tikzpicture}
	}} \right|^2  
\end{equation}
The OPE block in \eqref{threecopystate} is defined to reproduce this conformal block expansion; it acts within the tensor product of three Virasoro reps, as indicated by the three arrows in the diagram. 
The tensor network corresponding to this state obtained by truncating the sum to a window around the saddlepoint weights is given by, 
\begin{equation} \label{threecopyTN}
    \ket{\Psi}_*=\vcenter{\hbox{
\begin{tikzpicture}[scale=0.75]
\draw (4.3,0) -- (5.8,0);
\draw (4.3,0) -- (4.3+0.5,0.886);
\draw (4.3,0) -- (4.3+0.5,-0.886);
\draw (4.3+0.5,0.886) -- (5.8,0.886);
\draw (4.3+0.5,-0.886) -- (5.8,-0.886);
\draw[black,fill=ctensorcolor] (4.3,0) circle (0.3);
\node[above] at (5.3,-1.066) {$h_3^*$};
\node[above] at (5.3,0.666) {$h_1^*$};
\node[above] at (5.3,-0.2) {$h_2^*$};
\draw[black,fill=btensorcolor] (5.8,0.6) rectangle (7,1.2);
\draw[black,fill=btensorcolor] (5.8,-0.3) rectangle (7,0.3);
\draw[black,fill=btensorcolor] (5.8,-0.6) rectangle (7,-1.2);
\draw (7,0.85) -- (8,0.85);
\draw (7,0.95) -- (8,0.95);
\draw (7,0.05) -- (8,0.05);
\draw (7,-0.05) -- (8,-0.05);
\draw (7,-0.95) -- (8,-0.95);
\draw (7,-0.85) -- (8,-0.85);
\node[right] at (5.7,0) {$|\mathcal{B}_2|^2_*$};
\node[right] at (5.7,0.9) {$|\mathcal{B}_1|^2_*$};
\node[right] at (5.7,-0.9) {$|\mathcal{B}_3|^2_*$};
\end{tikzpicture}
}}
\end{equation}
Here, the saddlepoint OPE block $|\mathcal{B}|^2_*$ defines a map: $\mathcal{H}_{\text{prim}}\otimes\mathcal{H}_{\text{prim}}\otimes \mathcal{H}_{\text{prim}} \to \mathcal{H}_{\text{CFT}}\otimes\mathcal{H}_{\text{CFT}}\otimes \mathcal{H}_{\text{CFT}} $.  It factorises between the three boundaries into $|\mathcal{B}|^2_*=|\mathcal{B}_1|^2_*|\mathcal{B}_2|^2_*|\mathcal{B}_3|^2_*$ which follows from an argument similar to the one presented above for the two-boundary case. The above tensor network discretizes the three boundary spatial wormhole discussed in \cite{Balasubramanian:2014hda} (see also \cite{Peach:2017npp}), whose $t=0$ slice is topologically a pair of pants,
\begin{align}
    \vcenter{\hbox{
\begin{overpic}[width=1.8in,grid=false]{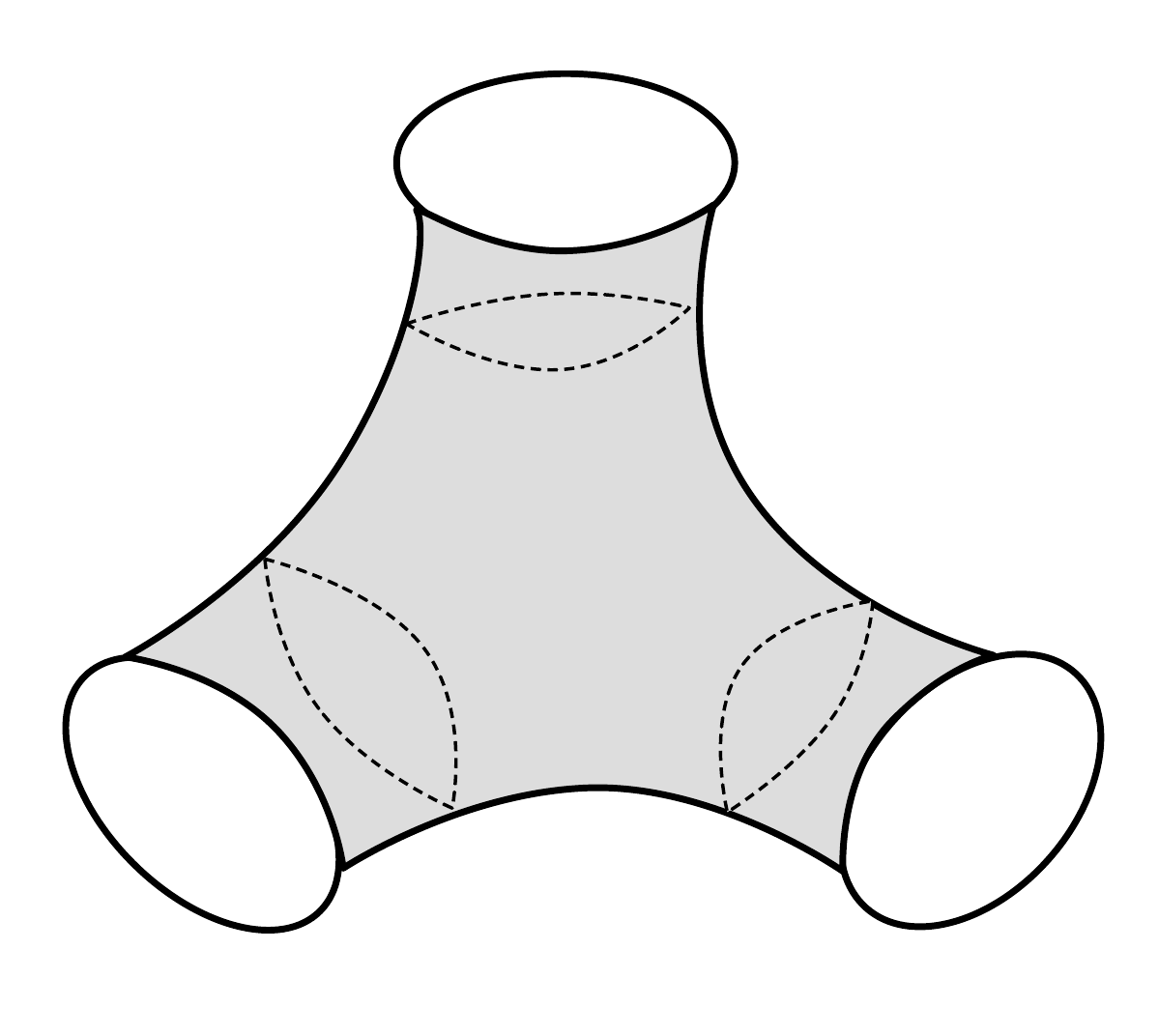}
\end{overpic}
}}
\end{align}
The lengths of the three geodesics are related to the saddlepoint weights by the relation $h_*=\frac{c}{24}(1+(\frac{\ell}{2\pi})^2)$. We can add EFT legs to the above network by adding probe matter to this background.

\section{Bulk geometries and the isometric transition}\label{s:isometric}

In this section, we describe the dual bulk geometries in detail, show that the bond dimensions in the tensor network agree with the areas of extremal surfaces, and check the isometric transition at the horizon for probe particles propagating on a large class of Euclidean black hole geometries in 3D. We must show that probe particles behind an apparent horizon act to \textit{decrease} the primary entropy in the OPE, while probe particles outside an apparent horizon act to \textit{increase} the primary entropy. 

We start by describing the construction of these black hole solutions by taking quotients of the three-dimensional hyperbolic space $\mathbb{H}_3$ by SL$(2,\mathbb{R})$ elements. Then, we take a short mathematical detour where we discuss a useful parametrisation of SL$(2,\mathbb{R})$ elements using which we shall derive mathematical identities involving the traces of these elements. We then provide a CFT interpretation for these identities which when combined with the ETH ansatz provides a derivation of the isometry properties of probes.

\subsection{Construction of black hole geometries}

The action of 3D gravity coupled to massive point particles is
\begin{align}\label{gravityaction}
S = -\frac{1}{16\pi G} \int_M \sqrt{g}(R+2)  - \frac{1}{8\pi G} \int \sqrt{h}(K-1) + \sum_i m_i \int dl_i \ ,
\end{align}
where the last integral is over the particle worldlines. The parameter $m_i$, with $0 < m_i < \frac{1}{4G}$, is referred to as the local mass of a particle; due to backreaction, it is not equal to the physical ADM mass. The ADM mass of a particle, or equivalently the total scaling dimension of the dual CFT operator, is 
\begin{align}
\Delta_i = m_i(1-2G m_i) \ .
\end{align}
Therefore with conformal weights parameterized as $h = \frac{c}{6}\eta(1-\eta)$, we have $m_i = \frac{\eta_i}{2G}$.
The point particles backreact on the geometry to produce conical defects of total angle $2\pi(1-2\eta_i)$. Since there are no propagating gravitons in the bulk, the set of solutions to (\ref{gravityaction}) can be classified in terms of smooth hyperbolic 3-manifolds, hyperbolic orbifolds, and similar quotients by elements of infinite order. We are interested in those solutions which can be interpreted as Euclidean black hole geometries. We choose a hyperbolic slicing, parametrising the metric on these geometries as
\begin{equation} \label{BHMetric}
    ds^2=d\tau^2+\cosh^2(\tau)d\Sigma^2
\end{equation}
where $\Sigma$ is a hyperbolic Riemann surface of constant negative curvature with one or more boundaries and/or conical defects. 
We assume the matter sources and boundary topology are such that $\Sigma$ admits a closed geodesic in the hyperbolic metric. These solutions are black holes --- the closed geodesic is the apparent horizon on the $t=0$ spatial slice. These coordinates can be analytically continued ($\tau\to it$) to FRW-like coordinates with the metric being,
\begin{equation}
    ds^2=-dt^2+\cos^2 t d\Sigma^2
\end{equation}
These coordinates cover the domain of dependence of the $t=0$ slice on the corresponding Lorentzian black hole geometry.

If all of the conical defects have finite order, then the spatial slice $\Sigma$ can be constructed as a quotient of the upper half plane $\mathbb{H}_2$ by a subgroup $\Gamma$ of SL$(2,\mathbb{R})$. The resulting 3-manifold is a quotient $\mathbb{H}_3/\Gamma$, with $\Gamma$ treated as a subgroup of the isometry group SL$(2,\mathbb{C})$ of $\mathbb{H}_3$. More generally, with conical defects of infinite order, the universal cover of $\Sigma$ is no longer $\mathbb{H}_2$ but it can be constructed similarly by identifying points of $\mathbb{H}_2$ under the action of  SL$(2,\mathbb{R})$ group elements. 
Precisely speaking, $\Sigma$ is uniformised by a conformal map to the upper half plane with its metric being the Liouville metric obtained by a pull-back of the Poincare metric on the upper half plane under the uniformisation map,
\begin{equation}\label{h2}
d\Sigma^2 = e^{\Phi}|dz|^2 = \frac{dy^2 + dx^2}{y^2}
\end{equation}
where $\Phi(z,\overline{z})$ is the Liouville field on $\Sigma$ and ($x,y)$ are the uniformising coordinates. To show that the metric (\ref{BHMetric}) agrees with the Poincare metric on $\mathbb{H}_3$, we make the following change of coordinates,
\begin{equation}
{\tilde y}= y \cos \phi,~~~~u= y \sin \phi,~~~~~\phi\equiv\cos^{-1}\left({\tanh}\tau\right)  
\end{equation}
the metric (\ref{BHMetric}) becomes
\begin{equation}\label{h3}
ds^2 = \frac{du^2+d{\tilde y}^2+dx^2}{u^2}~.
\end{equation}
This is the usual hyperbolic metric on ${\mathbb H}_3$, represented as the upper half 3-space with $u>0$. The full black hole geometry can be thought of as the surface of rotation about the $y=0$ axis, with the boundaries at $\tau=\pm \infty$ now identified as the surfaces $\phi=0,\pi$ where $u=0$. Note that if $\Sigma$ was a compact boundaryless Riemann surface, then the quotient construction would describe the Maldacena-Maoz wormhole \cite{Maldacena:2004rf}. Since we are interested in describing black hole geometries, we require $\Sigma$ to have one or more boundaries. In this case, the Liouville field $\Phi(z,\overline{z})$ solves the Liouville equation possibly in the presence of defects, with ZZ boundary conditions provided on each of the boundaries. The boundary components at $\tau=+\infty$ and $\tau =-\infty$ are glued together at the boundary of $\Sigma$, so that the conformal boundary is connected.

For example, the Liouville field corresponding to one-sided black hole geometries formed by the backreaction of two or more conical defects is determined by
\begin{equation}
  \partial\overline{\partial}\Phi=\frac{e^{\Phi}}{2}-2\pi\sum_i\eta_i\delta^{(2)}(z-z_i)
\end{equation}
subject to the boundary conditions,
\begin{equation}
    \Phi(z,\overline{z}) \sim
    \begin{cases}
      -4\eta_i\log(|z-z_i|) \quad & z\to z_i \\
      -2\log(1-|z|^2) \quad & |z|\to 1
    \end{cases}
\end{equation}
	Here, $z$ is a complex coordinate on the punctured unit disk with the defects located at $\{z_i\}$. The geometry of $\Sigma$ is illustrated in figure \ref{fig:introexample} in the introduction. More generally we can also include handles in $\Sigma$ so that the black hole has nontrivial topology behind the horizon. Such higher topology microstate geometries were discussed for instance in \cite{Maloney:2015ina}.

The Brown-York stress tensor obtained from \eqref{BHMetric} at $\tau = \pm \infty$ is equal to the semiclassical stress tensor of the auxiliary Liouville field,
\begin{align}
T(z) =  \frac{1}{2}  \p^2 \Phi - \frac{1}{4}(\p \Phi)^2   \ .
\end{align}
This can be used to find the primaries running in the dual OPE. The well known procedure (see \cite{ZamoRecursion} and for pedagogical discussions \cite{Harlow:2011ny,Hartman:2013mia}) is to study the monodromies of the Fuchsian differential equation,
\begin{align}\label{fuchseq}
\psi''(z) + \frac{6}{c}T(z) \psi(z) = 0  \ .
\end{align}
\newcommand{\bw}{\bar{w}}
This is equivalent to the Liouville equation, with $\Phi$ determined by the two solutions $\psi_1$, $\psi_2$ of this second order equation, 
\begin{align}
\Phi = \log \frac{4w'(z) \bw'(\bz)}{(1 - w(z) \bw(\bz))^2} , \quad w(z) = \frac{\psi_1(z)}{\psi_2(z)} \ . 
\end{align}
The condition that $\Phi$ is single-valued implies that around a closed loop $\gamma$ in the $z$ plane, the vector $\begin{pmatrix}\psi_1 \\ \psi_2\end{pmatrix}$ has monodromy $M(\gamma) \in {\rm SL}(2,\mathbb{R})$. The conformal weight of the primary in the semiclassical conformal block cut along the curve $\gamma$ is related to the monodromy by 
\begin{align}\label{trMrel}
\Tr M(\gamma) = 2 \cos(2\pi \eta)
\end{align}
with $h = \bh = \frac{c}{6}\eta(1-\eta)$.

\subsection{SL$(2,\mathbb{R})$ elements and monodromies}

Consider a general element of the SL(2,$\mathbb{R}$) group represented using a real $2\times 2$ matrix with unit determinant,
\begin{align}
    M=
    \begin{bmatrix}
    a & b\\
    c & d
    \end{bmatrix}
\quad \quad ad-bc=1
\end{align}
It generates an automorphism, $M: \mathbb{H}_2\to \mathbb{H}_2$ of the upper half plane by mapping points on the upper half plane by fractional linear transformations,
\begin{equation}
    z \to \frac{az+b}{cz+d} \quad \ , \quad z \in \mathbb{H}_2
\end{equation}
Depending on the conjugacy class of SL$(2,\mathbb{R})$ that the element $M$ belongs to, its action on $\mathbb{H}_2$ has fixed points either in the interior of $\mathbb{H}_2$ or on its boundary, i.e, on the real line. Let us denote the fixed point(s) of this map to be at $z=w_{1,2}$ which are roots of the quadratic equation,
\begin{equation}
    cw^2+(d-a)w-b=0
\end{equation}
Given the fixed points, there is a 1-parameter family of SL(2,$\mathbb{R}$) elements labelled by the entry $c$ below the diagonal without loss of generality. We can express the other entries of $M$ as
\begin{equation} \label{1parsoln}
    \begin{split}
        & b=-cw_1w_2 \\
        & a=\frac{c}{2}(w_1+w_2)\pm \frac{1}{2}\sqrt{4+c^2(w_1-w_2)^2}\\
        & d=-\frac{c}{2}(w_1+w_2)\pm \frac{1}{2}\sqrt{4+c^2(w_1-w_2)^2}
    \end{split}
\end{equation}
Using (\ref{1parsoln}), we see that
\begin{equation}
   \text{Tr}(M)=a+d=\pm \sqrt{4+c^2(w_1-w_2)^2}
\end{equation}
Depending on the conjugacy class that $M$ belongs to, we have
\begin{equation}
    \begin{split}
       \text{Elliptic}\, (|\text{Tr}(M_e)|<2): \quad & w_1=w_2^* \in \mathbb{H}_2\\
       \text{Parabolic}\, (|\text{Tr}(M_p)|=2): \quad & w_1=w_2 \in \mathbb{R} \quad \rm for \quad c\neq0\\
       \text{Hyperbolic}\, (|\text{Tr}(M_h)|>2): \quad & w_1 \neq w_2 \in \mathbb{R} 
    \end{split}
\end{equation}
Therefore, each elliptic element has a single fixed point in the interior of $\mathbb{H}_2$, each parabolic element has a single fixed point on the real line whereas each hyperbolic element has two fixed points on the real line.

Consider a hyperbolic 2-orbifold $\Sigma$, as in \eqref{h2}, constructed by  identifying $\mathbb{H}_2$ under the action of one or more SL$(2,\mathbb{R})$ elements. Each homology class $\gamma$ of simple closed curves on $\Sigma$ (with elliptic fixed points removed) corresponds to a conjugacy class in SL$(2,\mathbb{R})$, and is assigned a conformal weight by \eqref{trMrel}.
For example, if $\Sigma$ is a disk with $n$ defects and no handles, then the geometry \eqref{BHMetric} is dual to a $2n$-point function, and the saddlepoint weights appearing in \eqref{saddleweights} are determined by \eqref{trMrel} with the weight $h_k$ corresponding to a curve $\gamma_k$ encircling the defects $i_1, i_2, \dots, i_k
$, and $\Tr (M(\gamma_k)) = -2\cos\left( \pi \sqrt{1-24 h_k/c}\right)$.

A curve that can be deformed to a small circle around a defect operator with $\eta \in (0, \frac{1}{2})$ corresponds to an elliptic element of SL$(2,\mathbb{R})$. The overall sign in \eqref{trMrel} is a convention, because it is really PSL$(2,\mathbb{R})$ that acts on the upper half plane. With this sign convention, the defect operators above the multi-twist threshold of $\eta=\frac{1}{4}$ have $\text{Tr}(M_e)\in (-2,0)$ with the defect operator just below the black hole threshold ($\eta= \frac{1}{2}^-$) having $\text{Tr}(M_e)=-2^+$. For the elliptic element $M_e$ to correspond to the monodromy matrix of a sub-threshold operator, we require
\begin{equation}
    \pm\sqrt{1-c^2r^2\sin^2(\theta)}=\cos(2\pi\eta) \quad \equiv \quad |c|r\sin(\theta)=\sin(2\pi\eta)
\end{equation}
where the fixed points have been parameterized as $w_1 = re^{i\theta}$, $w_2 = r e^{-i\theta}$ with $r > 0$ and $\theta \in (0, \pi)$. 
This shows that we need to choose the positive branch for defects below the multi-twist threshold and the negative branch for defects above the multi-twist threshold. 
With this parametrisation, the family of elliptic elements in matrix form read
\begin{align} \label{ellipticparam}
M_e =
    \begin{bmatrix}
       \sin(2\pi\eta)\cot\theta+\cos(2\pi\eta) & -\sin(2\pi\eta)r\csc\theta \\
       \sin(2\pi\eta)\frac{\csc\theta}{r} & -\sin(2\pi\eta)\cot\theta+\cos(2\pi\eta)
    \end{bmatrix}
\end{align}
If the above monodromy matrix corresponds to a probe operator ($0 < \eta \ll 1$), we may express $M_e$ as a perturbation away from the identity,
\begin{align} \label{probeparam}
M_e = I+2\pi\eta
    \begin{bmatrix}
       \cot\theta & -r\csc\theta \\
       \frac{\csc\theta}{r} & -\cot\theta
    \end{bmatrix}+O(\eta^2)
\end{align}
Analytically continuing (\ref{trMrel}) to above the BH threshold using $\eta=\frac{1}{2}(1+i\lambda)$ with $\lambda \in \mathbb{R}^+$ corresponding to an operator with $h=\frac{c}{24}(1+\lambda^2)$, we have,
\begin{equation}
    \text{Tr}(M_h)=-2\cosh(\pi \lambda)
\end{equation}
so $\text{Tr}(M_h)<-2$. We also observe that heavier operators have monodromy matrices with smaller trace, i.e, 
\begin{equation}
    h'>h \implies \text{Tr}(M')<\text{Tr}(M)
\end{equation}
Note that the length of the primitive geodesic $\ell$ associated with the hyperbolic element $M_h$ which satisfies $\text{Tr}(M_h)=-2\cosh(\frac{\ell}{2})$ can now be expressed in terms of the chiral dimension $h$ of the primary operator running in an intermediate channel in a Virasoro OPE block as
\begin{equation}\label{hlength}
    \ell=2\pi\lambda \implies h=\frac{c}{24}\left(1+\left(\frac{\ell}{2\pi}\right)^2 \right)
\end{equation}
This equation confirms a claim made in section \ref{s:rtncft}: The bond dimensions in the tensor network agree with the areas of extremal surfaces in the bulk, since $\ell$ is the length of the extremal surface, and \eqref{hlength} is identical to the formula for the Cardy entropy with $S = \ell/4G$.

We can reparametrise the family of hyperbolic elements with fixed points at $w_1,w_2 \in \mathbb{R}$ and $w_1 < w_2$ in terms of the length of the primitive geodesic using  
$c=\frac{2\sinh(\frac{\ell}{2})}{w_1-w_2}$, which yields
\begin{align} \label{hyperbolicparam}
M_h =
    \begin{bmatrix}
       \sinh(\frac{\ell}{2})\frac{w_1+w_2}{w_1-w_2}-\cosh(\frac{\ell}{2}) & -2\sinh(\frac{\ell}{2})\frac{w_1w_2}{w_1-w_2}  \\ \frac{2\sinh(\frac{\ell}{2})}{w_1-w_2}
        & -\sinh(\frac{\ell}{2})\frac{w_1+w_2}{w_1-w_2}-\cosh(\frac{\ell}{2})
    \end{bmatrix} \ .
\end{align}

\subsection{Probes in pure state black holes}

We now turn to understanding the isometric property of probes for pure state black holes formed by backreaction of scalar defects and handles in the interior. The $t=0$ spatial slice for these black holes has a bulge in the interior, separated by one or more locally minimal surfaces from the asymptotic region.  The outermost minimal surface is the apparent horizon. These geometries are  examples of one-sided pythons \cite{Brown:2019rox}, with the `lunch' consisting of the region inside the apparent horizon. The product of operators creating the black hole will be called $\Psi = \O_1\O_2\O_3\cdots$, and the probe operator will be called $\O$, having weight $\eta \ll 1$. 

Denote the simple closed curve around the background operators alone by $\gamma_\Psi$, and the simple closed curve around all operators including the probe by $\gamma_{\Psi \O}$. The monodromies $M_i = M(\gamma_i)$ around these curves are related to the saddlepoint primary weights weights in the OPE, denoted $h_\Psi$ and $h_{\Psi \O}$ respectively and both above the black hole threshold, by \eqref{trMrel}, i.e., 
\begin{align}
\Tr M_\Psi &= 2 \cos 2\pi \eta_\Psi \\
\Tr M_{\Psi \O}  &= 2 \cos 2\pi \eta_{\Psi \O}
\end{align}
with $h_i = \frac{c}{6}\eta_i(1-\eta_i)$. Our goal is to compare $h_\Psi$ to $h_{\Psi \O}$. As explained in the previous section, if we assume eigenstate thermalization, then the code is isometric for $h_\Psi <h_{\Psi \O}$, and coisometric for $h_\Psi >h_{\Psi \O}$.

\begin{figure}
\begin{center}
\begin{overpic}[width=5.7in,grid=false]{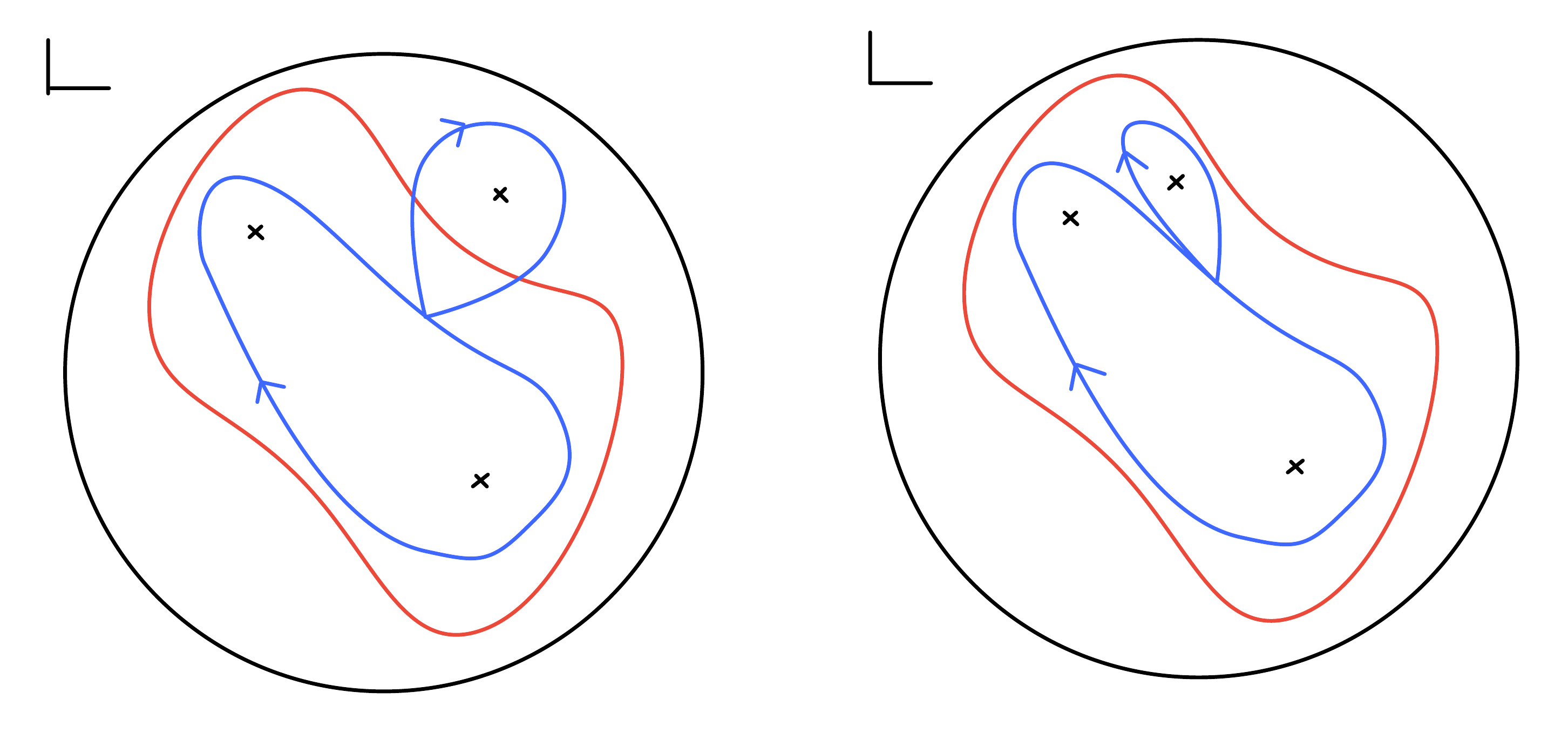}
\put (7,20) {$\gamma_{BH}$}
\put (15,29) {$\mathcal{O}_1$}
\put (26,15) {$\mathcal{O}_2$}
\put (29,36) {$\mathcal{O}_3$}
\put (32,20) {$\gamma_{\Psi}$}
\put (36,35) {$\gamma_e$}
\put (4,42) {$z$}
\put (37,5) {$|z|=1$}
\put (57,42) {$z$}
\put (61,20) {$\gamma_{BH}$}
\put (67,30) {$\mathcal{O}_1$}
\put (78,15) {$\mathcal{O}_2$}
\put (73,36) {$\mathcal{O}_3$}
\put (90,5) {$|z|=1$}
\put (84,21) {$\gamma_{\Psi}$}
\put (78,32) {$\gamma_e$}
\end{overpic}
\end{center}
\caption{
Spatial slice $\Sigma$ for a black hole created by $\Psi = \O_1(x_1) \O_2(x_2)$, probed by a third operator $\O_3(x_3)$. Monodromies are calculated around the blue curves, $\gamma_\Psi$ and $\gamma_{\Psi \O} = \gamma_\Psi \circ \gamma_{e}$, and the red curve $\gamma_{BH}$ is the apparent horizon. On the left, the probe is outside the horizon which is homologous to $\gamma_\Psi$, while on the right, the defect is inside  so the horizon is homologous to $\gamma_{\Psi \O}$.
\label{fig:zthree}
}
\end{figure}
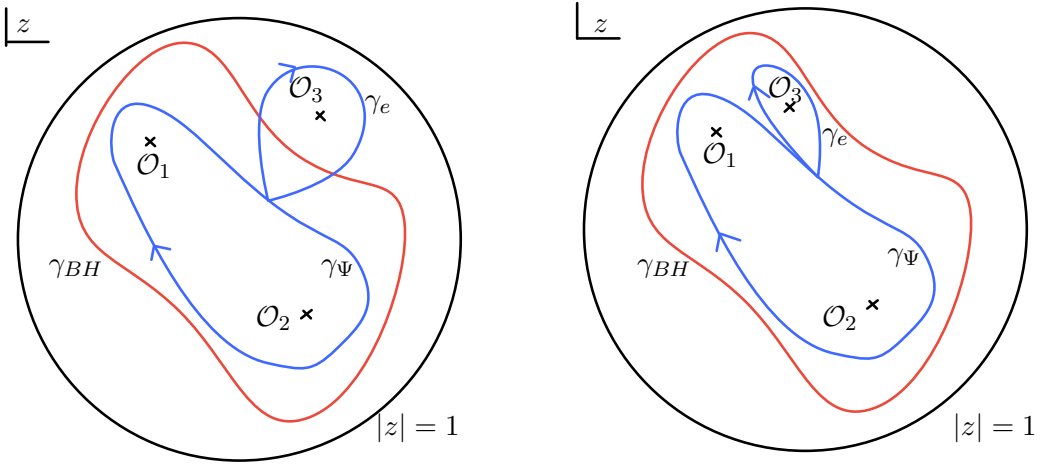

Although the analysis does not depend on the details of the black hole, an example to have in mind is a pure-state black hole created by two heavy defects, $\Psi = \O_1(x_1) \O_2(x_2)$, where $\eta_1, \eta_2 \in (\frac{1}{4}, \frac{1}{2})$. The background geometry is pictured in \eqref{spiky2}. The spatial geometry $\Sigma$ is found by solving the Liouville equation inside the unit disk with three defects, as shown in figure \ref{fig:zthree}. In this example there is a single minimal surface, which is a geodesic in the hyperbolic metric on $\Sigma$. The defect, $\O$, may be inside or outside this geodesic. The saddlepoint weights in the tensor network studied in section \ref{ss:probe6} were denoted there $h_1^* = h_{\Psi}$ and $h_2^* = h_{\Psi O}$.

\begin{figure}
\begin{center}
\begin{overpic}[width=5in]{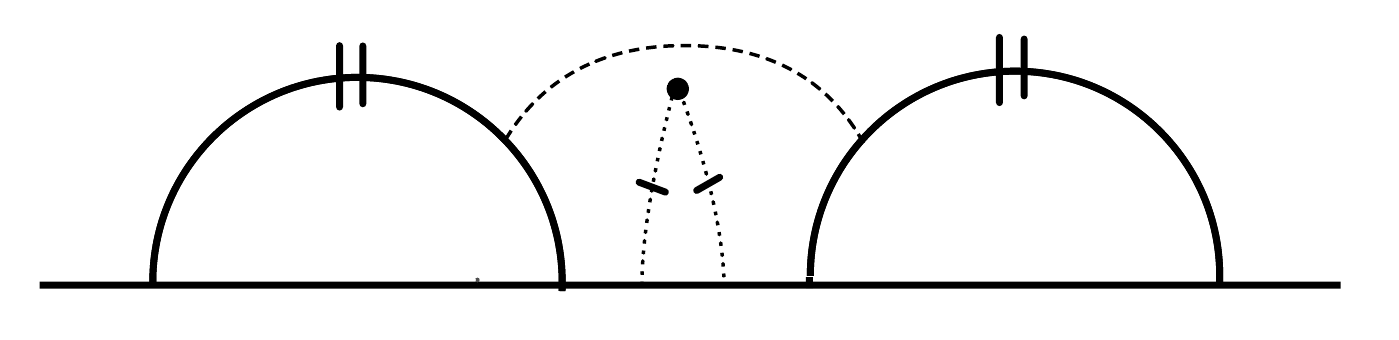}
\end{overpic}
\end{center}
\caption{Probe particle outside the apparent horizon of a black hole. The bold semicircles are identified by the action of the hyperbolic element $M_\Psi$ in \eqref{hypele}. The primitive geodesic associated with (\ref{hypele}) is represented by the dashed arc and is interpreted as the outermost horizon on the spatial slice of the pure state black hole geometry. The defect produces a small conical deficit by identifying the two dotted segments. There are assumed to be additional identifications behind the horizon, not shown, so that the boundary is $S^1$.
\label{fig:semicircles}}
\end{figure}

For the discussion of the isometric property we are interested in a probe that is outside, or just slightly inside, the apparent horizon. We can therefore focus on the part of the geometry that extends from the conformal boundary to slightly inside the horizon. After uniformizing to the $w$-plane, this part of the geometry is a quotient of $\mathbb{H}_2$. The case where the probe is outside the horizon is shown in figure \ref{fig:semicircles}. The identifications are generated by one elliptic element, corresponding to the probe, and one hyperbolic element, corresponding to the loop $\gamma_\Psi$ around the background operators. It is convenient to conjugate the hyperbolic element \eqref{hyperbolicparam} into the form
\begin{equation} \label{hypele}
    M_\Psi=\begin{bmatrix}
       -\cosh(\frac{\ell}{2}) & -\sinh(\frac{\ell}{2})\\
       -\sinh(\frac{\ell}{2}) & -\cosh(\frac{\ell}{2})
    \end{bmatrix}
\end{equation}
$M_\Psi$ has fixed points at $w=\pm 1$ and an associated primitive geodesic which is a portion of hyperbolic length $\ell$ of the semicircle centered at the origin with unit coordinate radius. We can choose a fundamental domain such that $M_\Psi$ identifies points on two semicircles related by reflection about $x=0$ on $\mathbb{H}_2$. This requirement uniquely determines the two semicircles to be centred at $(\pm x_0,0)$ each with radius $R$,
\begin{equation}
   x_0=\frac{1}{\tanh(\frac{\ell}{2})}, \quad \quad R=\frac{1}{\sinh(\frac{\ell}{2})}
\end{equation}
The two fixed points are on the real line, inside the semicircles, and with the choice \eqref{hypele} the point $w_1 = -1$ is repelling and $w_2 = +1$ is attractive.

The monodromy around the curve $\gamma_e$ circling the defect is elliptic, so it has one fixed point in the upper half plane at $w = re^{i\theta}$. This fixed point is the location of the probe particle in the uniformizing coordinate, and the monodromy matrix to leading order in the defect weight is given in \eqref{probeparam}. 

Using \eqref{hypele} and \eqref{probeparam}, observe the trace relation
\begin{align}\label{trgg}
\Tr M_{\Psi \O} = \Tr M_\Psi M_e  = \Tr M_\Psi   
-2\pi\eta \sinh(\frac{\ell}{2})\frac{(1-r^2)}{r\sin\theta}+O(\eta^2)
\end{align}
where $\eta$ corresponds to the probe weight.  In terms of the saddlepoint weights, this implies
\begin{align}
h_{\Psi O} - h_\Psi = \frac{m\ell}{8\pi}\frac{(1-r^2)}{r\sin\theta}
\end{align}
in the probe limit.
Therefore there is a transition from an isometric to a co-isometric code as the defect crosses the horizon at $r=1$. For $r<1$, the defect is outside the horizon,  $h_{\Psi \O} > h_{\Psi}$, and the code is isometric; for $r>1$, the defect is inside the horizon, the sign changes so that $h_{\Psi \O} < h_\Psi$, and the code is co-isometric.  

The primary weight $h_{BH}$ corresponding to the black hole horizon equals either $h_{\Psi}$ or $h_{\Psi \O}$, depending on whether the probe is outside or inside the horizon. If the probe is outside the horizon, then the curve $\gamma_{\Psi}$ is homologous to the horizon, so $h_{BH} = h_\Psi$. If the probe is inside, then the curve $\gamma_{\Psi \O}$ is homologous to the horizon, so $h_{BH} = h_{\Psi \O}$. See figure \ref{fig:zthree}. The black hole weight is related to the horizon length by \eqref{hlength}.

\bigskip

The trace identity \eqref{trgg} can easily be generalized to $n$ probes,
\begin{equation} \label{multraceoneside}
    \text{Tr}(M_\Psi\prod_{i=1}^n M_{e_i})=\text{Tr}(M_\Psi)-2\pi \sinh(\frac{\ell}{2})\sum_{i=1}^n \eta_i\frac{(1-r_i^2)}{r_i\sin\theta_i}+O(\eta_i\eta_j)
\end{equation}
Thus each probe inside (outside) the horizon acts to decrease (increase) the primary energy in the OPE. The ordering of the probe operators is not important at linear order.

So far, we have restricted discussion to the outermost geodesic on the spatial slice of pure state black hole geometries. However, when the black hole is formed due to backreaction by $n>2$ conical defects (taken to be above the multi-twist threshold), there are multiple geodesics on $\Sigma$.  On the CFT side, we expand the OPE of the conical defect operators in a channel where we fix the monodromies around curves homologous to a nested set of $n-1$ such geodesics, each circling one additional defect. The scaling dimensions of the internal scalars in this channel are given by 
\begin{equation} \label{Deltal}
    \Delta_i=2h_i=\frac{c}{12}(1+(\frac{\ell_i}{2\pi})^2)
\end{equation}
where $\ell_i$ are the lengths of the geodesics in the hyperbolic metric induced on the spatial slice. We observe from (\ref{Deltal}) that we can assign a coarse grained entropy that satisfies the Cardy formula for each of these geodesics,
\begin{equation}
    S_0(h_i,h_i)=4\pi\sqrt{\frac{c}{6}(h_i-\frac{c}{24})}=\frac{c\ell_i}{6}
\end{equation}
The additional geodesics can be detected in the dual CFT by adding a probe operator, which satisfies analogous trace identities near each geodesic. In the comb OPE channel, the probe transitions from isometric to co-isometric as it crosses the geodesic.

\subsection{Isometric transition of heavy operators} \label{nonperp}

So far, in this section, we have studied the isometry properties of probe operators in a black hole background created by heavy defects. Now, we would like to understand the isometric properties of the heavy defect operators that form the background. Consider a black hole geometry with its outermost horizon of length $\ell$ described by the primitive geodesic of the hyperbolic element $M_h$ in (\ref{hypele}) on the uniformizing upper half plane. Let one of the defects constituting this background be described by the elliptic element in (\ref{ellipticparam}) (which we shall denote as $M_d$ in this section) of strength $\eta$ whose fixed point is at $w=re^{i\theta}$  chosen to be above the primitive geodesic (i.e, $r>1$) on the uniformising upper half plane. Note the trace identity,
\begin{equation}
  \text{Tr}(M_h)-\text{Tr}(M_hM_d^{-1})=2\cosh(\frac{\ell}{2})(\cos(2\pi\eta)-1)+\sinh(\frac{\ell}{2})\sin(2\pi\eta)(r-\frac{1}{r})\csc(\theta)
\end{equation}
In the above expression, the first term is always negative and since we are assuming that the defect is behind the horizon, the second term is always positive. So, there is a locus $(r_0(\theta))$ behind the horizon corresponding to $\text{Tr}(M_h)=\text{Tr}(M_hM_d^{-1})$,
\begin{equation} \label{nphor}
   r_0(\theta)=\alpha(\theta)+\sqrt{1+\alpha^2(\theta)}, \quad \quad \alpha(\theta)=\coth(\frac{\ell}{2})\tan(\pi\eta)\sin(\theta)
\end{equation}
If the heavy defect lies on this locus, then (assuming ETH) the dual CFT operator ${\cal O}_d$ acts unitarily on the auxiliary Hilbert space spanned by Virasoro primaries in a microcanonical window around the scalar primary of scaling dimension $\Delta=\frac{c}{12}(1+(\frac{\ell}{2\pi})^2)$. If the defect is behind this locus, we have $\text{Tr}(M_h)>\text{Tr}(M_hM_d^{-1})$ so the defect operator acts co-isometrically on the relevant portions of the CFT Hilbert space whereas if the defect is outside this locus but still behind the horizon, the defect operator acts isometrically. Therefore from the point of view of the reconstructibility of operator ${\cal O}_d$, the unitary locus behaves like a `non-perturbative horizon'. This notion of horizon is operator-dependent due to backreaction.  Geometrically, the unitary locus corresponds to a spatial slice where the outermost and next-to-outermost geodesic have the same length; the action of ${\cal O}_d$ is isometric if the outermost geodesic is longer.

\section{Probes in multi-boundary black holes}\label{s:btz}

It is straightforward to extend all of the results above, including the semiclassical tensor network, to multi-boundary black holes. We will focus on checking the isometric property for probes in a deformed two-boundary black hole. For spherically symmetric states, the isometric property at finite temperature was shown in \cite{VerlindeBanff,VerlindeInProgress,Verlinde:2022xkw}. 

The Euclidean BTZ geometry can be constructed from a quotient of $\mathbb{H}_3$ by a discrete Abelian group (isomorphic to $\mathbb{Z}$) generated by
\begin{align} \label{BTZgen}
    M_h=
    \begin{bmatrix}
       -e^{\frac{2\pi^2}{\beta}} & 0\\
       0 & -e^{-\frac{2\pi^2}{\beta}}\\
    \end{bmatrix}
\end{align}
where $\beta$ is the inverse temperature of the BTZ black hole which is related to its ADM mass by $M_{\text{BTZ}}=\frac{c\pi^2}{3\beta^2}$. Such a quotient construction defines a natural slicing of the solid torus by hyperbolic cylinders. The BTZ metric in these coordinates is
\begin{equation}
    ds^2=d\tau^2+\cosh^2 \tau d\Sigma^2
\end{equation}
where $d\Sigma^2$ is the hyperbolic metric on the cylinder. Solving the Liouville equation on the cylinder with ZZ boundary conditions at either end ($\text{Im}(z)=0,\frac{\beta}{2})$, one can check that \cite{Seiberg:1990eb}
\begin{equation}
  d\Sigma^2=e^{\Phi(z,\overline{z})}|dz|^2=\frac{(\frac{2\pi}{\beta})^2}{\sin^2(\frac{2\pi}{\beta}\text{Im}(z))}|dz|^2
\end{equation}
with $z\sim z+2\pi $ and $\text{Im}(z) \in (0,\frac{\beta}{2})$.

 The $t=0$ spatial slice of the BTZ geometry has a minimal geodesic at the centre, whose length in the hyperbolic metric ($\ell(\gamma)=\frac{4\pi^2}{\beta}$) gives the area of the horizon. Notice that this length can be read off from the SL$(2,\mathbb{R})$ generator using $\text{Tr}(M_h)=-2\cosh (\frac{\ell(\gamma)}{2})$. The spatial slice can be conformally mapped to $\mathbb{H}_2$ with its image being a half-annulus between $|w|=1$ and $|w|=e^{\frac{4\pi^2}{\beta}}$ in $\mathbb{H}_2$ bounded by the real line and the circular boundaries identified under the action of $M_h$ in (\ref{BTZgen}).

\subsection{Probe in BTZ}
Consider a probe particle propagating in the BTZ background described by (\ref{BTZgen}). There is no nontrivial check of the isometric property in this case, but it is a useful warmup calculation. The corresponding tensor network was described in section \ref{sss:btznetwork}. Using the form of the SL$(2,\mathbb{R})$ element with fixed point at $w=re^{i\theta}$ ($r \in (1,e^{\frac{4\pi^2}{\beta}})$, $\theta \in (0,\pi)$) corresponding to a probe particle, we observe the following trace identity, 
\begin{equation} \label{Trid}
    \text{Tr}(M_hM_e)=\text{Tr}(M_h)-4\pi\eta\cot\theta \sinh(\frac{2\pi^2}{\beta})+O(\eta^2)
\end{equation}
In contrast to the black hole geometries described in the previous subsection, the BTZ geometry is spherically symmetric. So, for the unperturbed BTZ black hole, it is easy to verify that the primary energy agrees with the ADM mass, i.e, $E_p=\frac{c}{12}(\frac{\ell}{2\pi})^2=M_{BTZ}$. However, with the addition of the probe particle, the geometry is no longer spherically symmetric so we expect there is a non-zero contribution from the boundary gravitons in the calculation of the total energy hence the primary energy is not expected to be match with the total energy in the presence of the probe particle. Just like in the discussion with pure state black holes, the difference in monodromies (\ref{Trid}) can be translated to a difference in saddlepoint primary energies,\footnote{In Schwarzschild-like coordinates where the BTZ metric takes the form $ds^2=(r^2-r_h^2)d\Tilde{\tau}^2+\frac{dr^2}{r^2-r_h^2}+r^2d\phi^2$, the energy difference in (\ref{deltaEbtz}) reads $\Delta E_p=m\sqrt{r^2-r_h^2}$ where $r_h=\frac{2\pi}{\beta}$. To derive this relation, it is useful to note the relation between Schwarzschild-like coordinates and wormhole-like coordinates,
\begin{equation}
  \begin{split}
     & y= \frac{r_h}{r}e^{r_h\phi}\sqrt{1+(\frac{r^2}{r_h^2}-1)\sin(r_h\Tilde{\tau})} \\
     & x=\sqrt{1-\frac{r_h^2}{r^2}}\cos(r_h\Tilde{\tau})e^{r_h\phi} \\
     & \sinh(\tau) = \frac{r}{r_h}\sqrt{1-\frac{r_h^2}{r^2}}\sin(r_h\Tilde{\tau})
   \end{split}
\end{equation}
}
\begin{equation}\label{deltaEbtz}
    \Delta E_p= \frac{2\pi m}{\beta} \cot(\theta)
\end{equation}
where $m$ is the mass of the probe.

The BTZ geometry with a probe inserted calculates a thermal 2-point function. Let us expand this 2-point function in Virasoro conformal blocks,
\begin{align} \label{blockBTZprobe}
\langle \O(z_1, \bz_1)\O(z_2, \bz_2) \rangle_{\tau,\btau} = \sum_{p,q} \left| c_{Opq} \right|^2 \left|\vcenter{\hbox{
	\begin{tikzpicture}[scale=0.75]
	\draw[thick] (0,0) circle (1);
	\draw[thick] (-1,0) -- (-2,0);
	\node[left] at (-2,0) {$\mathcal{O}$};
	\node[above] at (0,1) {$p$};
	\node[below] at (0,-1) {$q$};
	\draw[thick] (1,0) -- (2,0);
	\node[right] at (2,0) {$\mathcal{O}$};
	\end{tikzpicture}
	}} \right|^2
\end{align}
The loop represents the thermal circle, $z \sim z + \tau$. The two hyperbolic identifications in the BTZ+probe geometry calculate the saddlepoint primary weights in this conformal block expansion,
\begin{align}
\Tr M_h  = -2 \cos \left( \pi \sqrt{1-24h_p^*/c}\right) , \quad
\Tr M_h M_e = -2 \cos \left( \pi \sqrt{1-24h_q^*/c}\right) \ . 
\end{align}
Thus $\Delta E_p$ in \eqref{deltaEbtz} is the difference in primary energies between the two internal lines in the conformal block at the saddlepoint. The larger weight appears on whichever arc has less Euclidean time evolution. In the bulk, the probe particle can be inserted in either the left or right side of the Penrose diagram. The relation \eqref{deltaEbtz} simply says that the side with the particle has a higher primary energy.

Even when the probe particle in the above setup is replaced by a heavy defect, it is clear that the saddlepoint weights $h_p^*$ and $h_q^*$ in (\ref{blockBTZprobe}) match if the particle propagates through the middle of the backreacted geometry (i.e, when the Euclidean time difference between the endpoints of the trajectory is half the size of the thermal circle). In this case, as explained in some more detail in the next subsection, the two apparent horizons created on either side of the particle's trajectory have equal area ensuring that the saddlepoint weights in the conformal block expansion match. When the defect propagates asymmetrically, it increases the saddlepoint primary energy on the side with the defect.
Thus, the middle of the punctured cylinder constitutes a unitary locus for a defect of any strength and hence behaves like a non-perturbative horizon in the sense of section \ref{nonperp}. As explained in footnote \ref{footnotenp}, due to the chosen parametrisation of the SL$(2,\mathbb{R})$ elements in the definition of the background geometry, the locus for the non-perturbative `horizon' matches with (\ref{nphor}).

\subsection{Probe in the BTZ+Defect background} \label{PETSgeo}

In the previous subsection, the distinction between the primary and total energies for the background geometry were unimportant because the BTZ geometry is spherically symmetric. However, we can deform the BTZ background by a heavy particle which manifestly breaks the spherical symmetry. In this case, the spatial slice is a once-punctured hyperbolic cylinder shown in \eqref{btzdf}. This is an example of a two-sided python's lunch \cite{Brown:2019rox}. For discussion of such geometries in JT gravity, see \cite{Goel:2018ubv,Stanford:2020wkf,Saad:2019pqd,Lin:2022zxd}. Spherically symmetric two-sided pythons in higher dimensional Einstein gravity were constructed in \cite{Sasieta:2022ksu}.  

\begin{figure}
\begin{center}
\begin{overpic}[width=4in, grid=false]{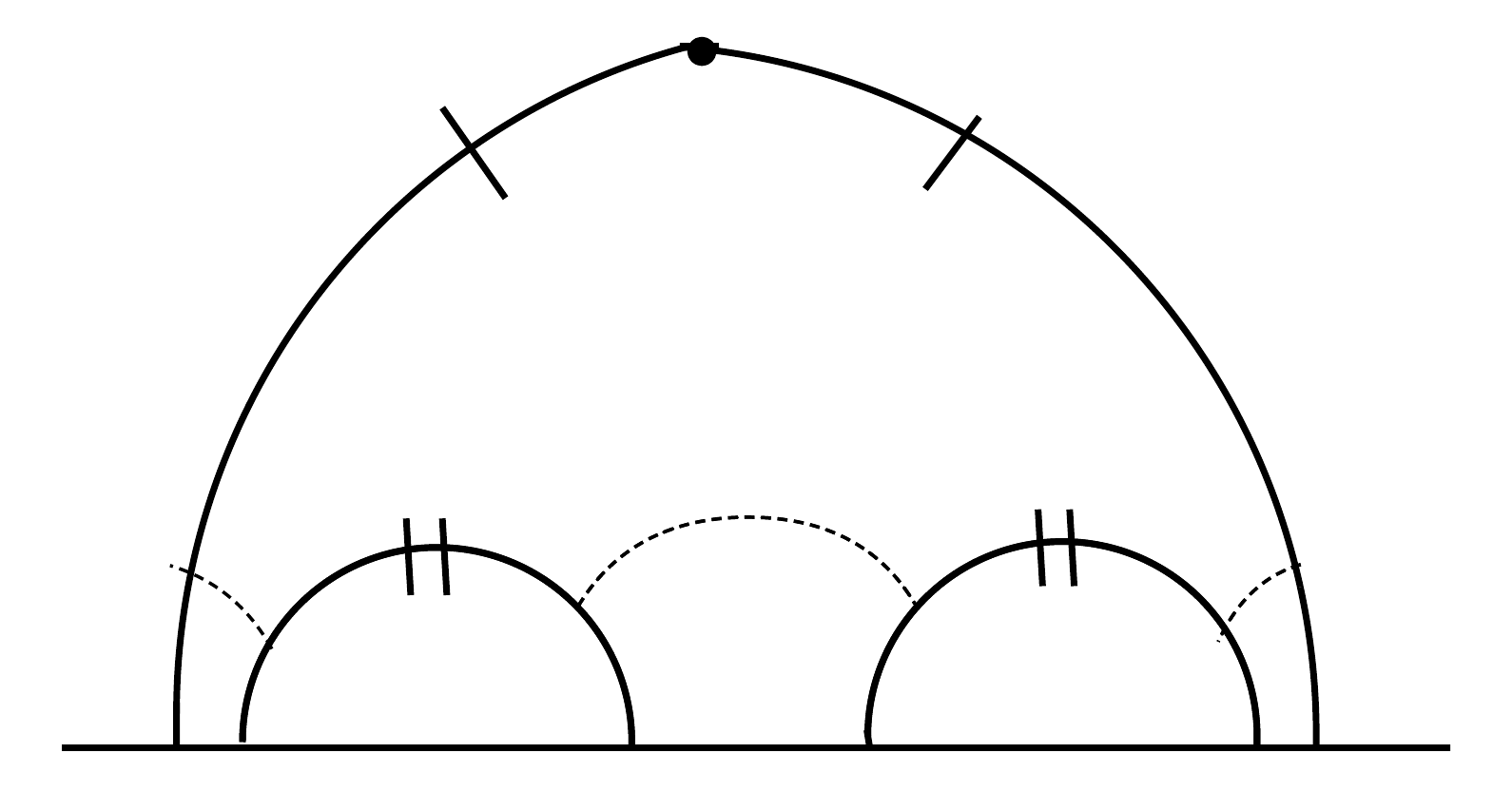}
\put (47,22) {$M_h$}
\put (15,15) {$M_b$}
\put (78,15) {$M_b$}
\end{overpic}
\end{center}
\caption{Quotient construction of the spatial slice of the BTZ+Defect black hole geometry. The dotted lines are the two apparent horizons which are primitive geodesics of hyperbolic elements $M_h$ and $M_b$.
 \label{fig:BTZ+Defect}}
\end{figure}

This BTZ+Defect geometry is constructed as follows. Let the elliptic element corresponding to the heavy particle be denoted $M_d$. Given the strength $\eta$ and the location of the fixed point $w=iy_0$ on $\mathbb{H}_2$, we can write down a matrix form for $M_d$,
\begin{equation}
    M_d=\begin{bmatrix}
       \cos(2\pi\eta) & y_0\sin(2\pi\eta) \\
       -y_0^{-1}\sin(2\pi\eta) & \cos(2\pi\eta)
    \end{bmatrix}
\end{equation}
There are two apparent horizons, one separating the lunch from each asymptotic region. Let one of them have length $\ell$. We can choose a canonical form of the hyperbolic element whose primitive geodesic has length $\ell$,
\begin{equation}
     M_h=\begin{bmatrix}
       -\cosh(\frac{\ell}{2}) & -\sinh(\frac{\ell}{2}) \\
       -\sinh(\frac{\ell}{2}) & -\cosh(\frac{\ell}{2})
    \end{bmatrix}
\end{equation}
The other apparent horizon corresponds to the primitive geodesic associated with the hyperbolic element $M_b=M_dM_h$. In matrix form,
\begin{equation} \label{hypelement}
    M_b=\begin{bmatrix}
       -\cos(2\pi\eta)\cosh(\frac{\ell}{2})-y_0\sin(2\pi\eta)\sinh(\frac{\ell}{2}) & -\cos(2\pi\eta)\sinh(\frac{\ell}{2})-y_0\sin(2\pi\eta)\cosh(\frac{\ell}{2}) \\
       y_0^{-1}\sin(2\pi\eta)\cosh(\frac{\ell}{2})-\cos(2\pi\eta)\sinh(\frac{\ell}{2}) & y_0^{-1}\sin(2\pi\eta)\sinh(\frac{\ell}{2})-\cos(2\pi\eta)\cosh(\frac{\ell}{2})
    \end{bmatrix}
\end{equation}
The parameters need to satisfy the constraint coming from the requirement that $M_b$ is hyperbolic, i.e, $\text{Tr}(M_b)<-2$,
\begin{equation}
    2\left(1-\cosh(\frac{\ell}{2})\cos(2\pi\eta)\right)<\sin(2\pi\eta)\sinh(\frac{\ell}{2})(y_0-y_0^{-1})
\end{equation}
If the heavy particle is placed in the middle of the space, then the lengths of the two apparent horizons are equal. For that case, we can relate the defect strength to the location of the fixed point on $\mathbb{H}_2$ using\footnote{We can use the general form of the defect elliptic element given in (\ref{ellipticparam}) with fixed point at $w=re^{i\theta}$ on $\mathbb{H}_2$ to determine the locus of fixed points corresponding to the middle of the space. Due to the form of the chosen elliptic and hyperbolic elements, this locus is given by (\ref{nphor}). \label{footnotenp}}
\begin{equation}
    \text{Tr}(M_b)=-2\cosh(\frac{\ell}{2}) \implies \cot(\pi\eta)=\frac{2\coth(\frac{\ell}{2})}{y_0-y_0^{-1}}
\end{equation}
The discussion of the isometry property around the apparent horizon corresponding to $M_h$ follows trivially from the discussion of the isometry property for pure state black holes due to the form of the chosen hyperbolic element, as the geometry is locally identical to figure \ref{fig:semicircles}. To verify the isometric property of probes around the other apparent horizon, note that,
\begin{multline}\label{defectTraceDiff}
    \text{Tr}(M_bM_e)-\text{Tr}(M_b)=-2\pi\eta_p\bigg[(y_0+y_0^{-1})\cot(\theta)\sin(2\pi\eta)\sinh(\frac{\ell}{2})\\+\csc(\theta)\sin(2\pi \eta)\cosh(\frac{\ell}{2})(\frac{y_0}{r}+\frac{r}{y_0})-\csc(\theta)\cos(2\pi\eta)\sinh(\frac{\ell}{2})(r-\frac{1}{r})\bigg]
\end{multline}
where $M_e$ is the elliptic element corresponding to the probe of strength $\eta_p$ having a fixed point at $w=re^{i\theta}$. The reader can verify that the locus $\text{Tr}(M_bM_e)=\text{Tr}(M_b)$ corresponds to the other apparent horizon by matching the matrix form of $M_b$ in (\ref{hypelement}) with the form of the hyperbolic element given by (\ref{hyperbolicparam}) to determine the fixed points of $M_b$ and then use the fact that the primitive geodesic is a portion of a semicircle which when extrapolated meets the real axis at the fixed points of the hyperbolic element.

In the dual CFT these results are interpreted in terms of the conformal block expansion for the finite-temperature 4-point function,
\begin{align}\label{ggha}
\langle \O_d \O_d \O_1 \O_1\rangle_{\tau,\btau} = 
\sum_{p,q,r,s} c_{sdp}c^*_{sdr}c_{p1q}c^*_{r1q} \left |
\vcenter{\hbox{
	\begin{tikzpicture}[scale=0.75]
	\draw[thick] (0,0) circle (1);
	\draw[thick] (-1,0) -- (-2,0);
	\node[left] at (-2,0) {$\mathcal{O}_d$};
	\node[above] at (0,1) {$s$};
	\node[below] at (0,-1) {$q$};
	\draw[thick] (1,0) -- (2,0);
        \draw[thick] (0.5,-0.866)--(0.5+0.5,-0.866-0.866);
        \draw[thick] (-0.5,-0.866)--(-0.5-0.5,-0.866-0.866);
        \node[right] at (0.866,-0.5) {$r$};
        \node[left] at (-0.866,-0.5) {$p$};
	\node[right] at (2,0) {$\mathcal{O}_d$};
        \node[below] at (0.5+0.5,-0.866-0.866) {$\mathcal{O}_1$};
        \node[below] at (-0.5-0.5,-0.866-0.866) {$\mathcal{O}_1$};
	\end{tikzpicture}
	}}  \right |^2
\end{align}
where $\O_d$ is the heavy particle corresponding to the elliptic element $M_d$, and $\O_1$ is the probe, which  are assumed to be inserted in a reflection-positive configuration.  The points $\tau=0,\frac{\beta}{2}$ that join onto the Lorentzian spacetime are on legs labeled $s$ and $q$, so the saddlepoint weights $h_s^*$ and $h_q^*$ are the (chiral) energies observed at infinity on the left and right sides of the black hole. The probe operator $\O_1$ is effectively a map 
\begin{align}
\O_1 : \H_p^* \to \H_q^* \ , 
\end{align}
with the direction corresponding to increasing Euclidean time. The question of whether the probe operator $\O_1$ acts isometrically is therefore answered by comparing the saddlepoint primary weights $h_q^*$ and $h_p^*$. 
The hyperbolic elements appearing in \eqref{defectTraceDiff} are related to the saddlepoint primary weights by
\begin{align}
\Tr M_b  = -2 \cos \left( \pi \sqrt{1-24h_p^*/c}\right) , \quad
\Tr M_b M_e^{-1} = -2 \cos \left( \pi \sqrt{1-24h_q^*/c}\right)  \ .
\end{align}
(Note $h_p^* = h_r^*$). Therefore, when the probe is outside the horizon, $h_q^* > h_p^*$, so the probe operator $O_1$ acts isometrically, and when it is inside the horizon, it acts co-isometrically.

\section{Discussion}

We have described two simple models where the CFT operator algebra can be recast as a pseudorandom tensor network: high-energy spherically symmetric states in arbitrary dimensions, and more general black hole states in a 2d CFT dual to pure gravity plus point particles. In both cases, the tensor network discretizes the radial direction in the bulk, in the sense that $(1)$ bond dimensions agree with the areas of extremal surfaces, and $(2)$ probes undergo an isometric transition at the horizon.

\bigskip
\noindent In the rest of this discussion we comment on several open directions.

\subsubsection*{Singularities}
The usual black hole singularity requires time evolution in Lorentzian signature. However, we can produce a very similar effect by moving the heavy operator insertions $\O_i$ toward the origin (in radial quantization) or large negative Euclidean time (on the cylinder). This results in a black hole at $t=0$ with a very narrow throat at the extremal surface, and therefore small entropy. By tuning the operator weights and Euclidean positions, we can send the horizon area toward zero. This is a Euclidean version of a black hole singularity. In this limit, the counting argument implies that the rank of $\O^\dagger \O$, with $\O$ dual to a probe particle behind the horizon, goes to zero. Therefore, in the singular limit, it becomes impossible to reconstruct behind-the-horizon operators from the boundary. 

What happens to our black hole solutions in this limit? The black hole geometry corresponds to the identity block approximation. The mass of the black hole is set by the saddlepoint weight $h^*$ in the dual OPE channel, and as we tune toward a singularity, this weight approaches the black hole threshold, $h^* \to \frac{c}{24}$. However, assuming there is any light matter in the theory, eventually the identity block approximation breaks down. In 3d gravity coupled to massive point particles, this breakdown can be studied quantitatively. There is an exchange of dominance between the black hole and the handle wormholes found in \cite[section 6]{Chandra:2022bqq}. If the black hole mass is very small, then instead of a black hole, the norm $\langle \Psi|\Psi\rangle$ is dominated by the handle wormhole. The $t=0$ slice, instead of being a smooth black hole, has two disconnected components --- the black hole interior is replaced by a closed universe with an additional heavy defect having $h^* < \frac{c}{24}$. That is, the outermost throat shown in figure \ref{fig:introexample} pinches off and breaks the diagram in two, with two new defect operators appearing at the singularities. It would be interesting to understand this regime better in terms of the holographic code.

\subsubsection*{Discretizing the transverse directions}
In the original formulation of holographic tensor networks using MERA \cite{Swingle:2009bg}, as well as the HaPPY code \cite{Pastawski:2015qua} and random tensor networks \cite{Hayden:2016cfa}, the spatial directions along the boundary are also discretized. In our model, each boundary component has only a single tensor. This allows for some simple tests of bulk reconstruction --- for example, it is easy to see from the isometric property that an operator inside a 2-sided black hole can be reconstructed from one boundary but that this does not hold in a 3-boundary black hole --- but it does not allow for spatial resolution on a single boundary. Can the CFT construction be generalized to write a more complete tensor network in terms of OPE data?

\subsubsection*{Crossing symmetry}

To construct the tensor network, we first chose an OPE channel. What if we choose a different channel? For example, in a case like figure \ref{fig:introexample} in the introduction, we could construct the comb channel OPE in a different order. This will give a different tensor network, but the two quantum states must agree.

This means that the OPE coefficients cannot be truly (psuedo)random: crossing symmetry requires corrections to the ETH ansatz \cite{Belin:2021ryy}. Throughout the paper we have assumed these corrections can be neglected. It would be interesting to explore how the isometric property is realized in other channels and when these corrections must be taken into account.

\subsubsection*{Higher dimensions: beyond spherical symmetry}
Can holographic tensor networks be constructed quantitatively in higher dimensions? As we have emphasized, this is a difficult problem without spherical symmetry, because dynamical light fields come into play (the same is true for 2d or 3d gravity with light matter). We can gain some rough intuition from the example of 2d CFT coupled to point particles. The theory must first be separated into `fast' and `slow' degrees of freedom: the black hole microstates and the low energy fields. Only the microstates can be approximated by random tensors, so the tensor network should consist of random tensors dressed by light fields. In some cases the distinction between the two is blurred by quantum scars \cite{Dodelson:2022eiz}. One proposal to build more realistic holographic tensor networks is to use random tensors with nontrivial links \cite{Cheng:2022ori}. However, in our model, upgrading the tensor network (figure \ref{fig:psinetwork}) to the exact CFT state \eqref{Bexp} does not appear to be as simple as weighting the tensors.

\subsubsection*{Coarse graining tensor networks and wormholes}

In \cite{Chandra:2022fwi} we showed that Euclidean wormholes calculate the replica partition functions of coarse grained states. The coarse graining operation involves projecting onto diagonal states in the bra and ket of a pure state $\rho = |\Psi\rangle\langle\Psi|$. The same procedure can be applied to tensor networks. Consider, for example, the tensor network in \eqref{threeopTN}, for which the density matrix is
\begin{align}
\rho &=\vcenter{\hbox{
\begin{tikzpicture}[scale=0.75]
\draw (1.5, -.05) -- (2.5,-.05);
\draw (1.5, .05) -- (2.5,.05);
\begin{scope}
\clip (2.5,-1) rectangle (3,1);
\draw[black,fill=btensorcolor] (2.5,0) circle (0.5);
\end{scope}
\draw (2.5,-0.5) -- (2.5,0.5);
\draw (3,0) -- (4,0);
\draw (4.3,0) -- (5.3,0);
\draw[black,fill=ctensorcolor] (4.3,0) circle (0.3);
\draw[black,fill=ctensorcolor] (5.6,0) circle (0.3);
\draw (-1.6+8.6,0) -- (-0.3+8.6,0);
\draw[black,fill=ctensorcolor] (-1.6+8.6,0) circle (0.3);
\draw[black,fill=ctensorcolor] (-0.3+8.6,0) circle (0.3);
\draw (0+8.6,0) -- (1+8.6, 0);
\draw (1.5+8.6, -.05) -- (2.5+8.6,-.05);
\draw (1.5+8.6, .05) -- (2.5+8.6,.05);
\draw[thick] (1.5+8.6,-0.5) -- (1.5+8.6,0.5);
\begin{scope}
\clip (1+8.6,-1) rectangle (1.5+8.6,1);
\draw[black,fill=btensorcolor] (1.5+8.6,0) circle (0.5);
\end{scope}
\end{tikzpicture}
}} 
\end{align}
Define the coarse-grained state by
\begin{align}\label{defRhoOuter}
{\cal C}(\rho) &= 
   \vcenter{\hbox{
\begin{tikzpicture}[scale=0.75]
\draw (1.5, -.05) -- (2.5,-.05);
\draw (1.5, .05) -- (2.5,.05);
\begin{scope}
\clip (2.5,-1) rectangle (3,1);
\draw[black,fill=btensorcolor] (2.5,0) circle (0.5);
\end{scope}
\draw (2.5,-0.5) -- (2.5,0.5);
\draw (3,0) -- (4,0);
\draw (4.3,0) -- (5.3,0);
\draw[black,fill=ctensorcolor] (4.3,0) circle (0.3);
\draw[black,fill=ctensorcolor] (5.6,0) circle (0.3);
\draw (-1.6+8.6,0) -- (-0.3+8.6,0);
\draw[black,fill=ctensorcolor] (-1.6+8.6,0) circle (0.3);
\draw[black,fill=ctensorcolor] (-0.3+8.6,0) circle (0.3);
\draw (0+8.6,0) -- (1+8.6, 0);
\draw (1.5+8.6, -.05) -- (2.5+8.6,-.05);
\draw (1.5+8.6, .05) -- (2.5+8.6,.05);
\draw[thick] (1.5+8.6,-0.5) -- (1.5+8.6,0.5);
\begin{scope}
\clip (1+8.6,-1) rectangle (1.5+8.6,1);
\draw[black,fill=btensorcolor] (1.5+8.6,0) circle (0.5);
\end{scope}
\draw (3.5,0) -- (3.5,1);
\draw (3.5,1) -- (9.1,1);
\draw (9.1,1) -- (9.1,0);
\draw[black,fill=black] (3.5,0) circle (0.1);
\draw[black,fill=black] (9.1,0) circle (0.1);
\end{tikzpicture}
}} 
\end{align}
where the black dot represents the diagonal 3-index tensor, $\delta_{pq}\delta_{qr}$. That is, we project onto identical Virasoro representations in the bra and ket. In the random tensor approximation, the coarse-grained entropy 
\begin{align}
S_{\rm coarse}(\rho) &:= -\tr {\cal C}(\rho) \log {\cal C}(\rho)
\end{align}
is equal to one quarter the area of the outermost apparent horizon. Furthermore, replicas such as $\Tr {\cal C}(\rho)^2$ have tensor networks that discretize multiboundary Euclidean wormholes in AdS$_3$. We hope to explore this  in future work.

\bigskip

\noindent \textbf{Acknowledgments} We thank Nima Afkhami-Jeddi, Ahmed Almheiri, Ven Chandrasekaran, Scott Collier, Tom Faulkner, Yikun Jiang, Juan Maldacena, Alex Maloney, Don Marolf, Sean McBride, Baur Mukhametzhanov, Onkar Parrikar, and Pratik Rath for helpful discussions. This work was supported by NSF grant PHY-2014071. Part of this work was done at the KITP, with support from NSF grant PHY-1748958.

\renewcommand{\baselinestretch}{1}\small
\bibliographystyle{ourbst}
\bibliography{biblio3}

\end{document}